\documentclass[useAMS,usenatbib]{mn2e}
\usepackage{graphicx}
\usepackage{times}

\begin{document}

\newcommand{\bfit}[1]{\mbox{\boldmath $#1$}}
\newcommand{\bfsf}[1]{\mbox{\boldmath \sf #1}}

\title[Ring-image turbulence sensor]{Measurement of turbulence profile 
from defocused ring images}

\author[Tokovinin]{A.~Tokovinin\thanks{E-mail:
atokovinin@ctio.noao.edu} \\
Cerro Tololo Inter-American Observatory / NSFs' NOIRlab, Casilla 603, La Serena, Chile\\
}

\date{-}

\pagerange{\pageref{firstpage}--\pageref{lastpage}} \pubyear{2020}

\maketitle

\label{firstpage}

\begin{abstract}
A defocused image of a bright single  star in a small telescope contains
rich information  on the  optical turbulence,  i.e.  the  seeing.  The
concept of a novel turbulence  monitor based on recording sequences of
ring-like intrafocal images  and their analysis is  presented.  It can
be   implemented    using   standard   inexpensive    telescopes   and
cameras. Statistics of  intensity fluctuations in the  rings and their
radial  motion  allow  measurement   of  the  low-resolution  turbulence
profile, the  total seeing,  and the  atmospheric time  constant.  The
algorithm  of  processing the  images  and  extracting the  turbulence
parameters   is  developed   and  extensively   tested  by   numerical
simulation.   Prescriptions to  correct for  finite exposure  time and
partially saturated  scintillation are given.  A  prototype instrument
with a 0.13-m aperture was tested  on the sky.  The RINGSS (Ring-Image
Next  Generation  Scintillation Sensor)  can  be  used as  a  portable
turbulence  monitor for  site testing  and as  an upgrade  of existing
seeing monitors.
\end{abstract}


\section{Introduction}

A live image  of a bright star  in a small slightly  defocused telescope
displays    distortions   caused    by    the   optical    turbulence,
``seeing''. Movies  of such images  captured by  a fast camera  can be
interpreted to measure turbulence strength, its distribution along the
line  of sight,  and  the characteristic  time  scale.  Extraction  of
quantitative  information  on  turbulence  from  the  fast  movies  of
ring-like defocused images is the  subject of this paper. A turbulence
monitor based  on this idea consists  of a small telescope  and a fast
camera;  its hardware  is standard  and inexpensive,  while the  major
challenge lies in the software needed to process the movies.

Measurement of the vertical distribution  of optical turbulence in the
terrestrial atmosphere (OTP --  optical turbulence profile) serves to
support operation  of modern astronomical observatories  equipped with
adaptive optics (AO) instruments  and to characterize new astronomical
sites.  A classical instrument to  measure the OTP using scintillation
of double stars, SCIDAR, needs aperture sizes on the order of 1\,m and
therefore   it   is   not    suitable   for   testing   remote   sites
\citep[e.g.][]{Klueckers1998}.  A  Multi-Aperture Scintillation Sensor
(MASS)  delivers low-resolution  OTPs  using  scintillation of  bright
single  stars  and  a  small aperture  of  $\sim$0.1\,m  \citep{MASS}.
Combination  of  MASS  with  the  Differential  Image  Motion  Monitor
\citep[DIMM,][]{DIMM} in one instrument  attached to a small telescope
has  become   a  standard  tool   for  site  testing   and  monitoring
\citep{Kornilov2007}.  About 35 such  instruments have been fabricated
and used in the site characterization projects \citep[e.g.][]{TMT} and
for turbulence monitoring at existing observatories.

MASS  records  scintillation  signals  from  four  concentric  annular
apertures  using photo-multipliers.   This technology  became obsolete
when     fast    low-noise     panoramic     detectors,    such     as
electron-multiplication   (EM)  CCDs   and  scientific   CMOS,  became
available.   Moreover,  the  opto-mechanics  and  electronics  of  the
MASS-DIMM  instruments are  custom-made  and  difficult to  replicate.
Nowadays,  MASS-DIMM should  be  replaced by  an  instrument based  on
solid-state light detectors.

The need to find an alternative to MASS has been generally recognized.
In a  master thesis  project, \citet{Kohlman} constructed  a prototype
where the flux  in four concentric annuli within  a 15-cm unobstructed
aperture was measured by a CMOS camera, emulating MASS. An alternative
approach is adopted by the  team at the Pontificia Catolica University
in Chile \citep{FASS}; they also  record intensity fluctuations at the
pupil,  but interpret  them  differently by  taking a  one-dimensional
Fourier transform in the angular coordinate of narrow rings carved from
the pupil images.  This approach  circumvents the problem of centrally
obstructed apertures of  standard small Schmidt-Cassegrain telescopes.
The instrument  is called  FASS (Full-Aperture  Scintillation Sensor),
and its development continues \citep{FASS2}.

Recording  scintillation   signals  at  the  telescope   pupil  with  a
solid-state detector requires a  large optical de-magnification factor
$k_{\rm magn}  \sim 10^3$ to match  the small physical size  of pixels
and the  spatial scale  of scintillation.   The latter  is set  by the
Fresnel radius $\sqrt{\lambda z} = 1.7$\,cm for a propagation distance
of $z = 0.5$\,km and a  wavelength of $\lambda = 0.6$\,$\mu$m.  If the
detector pixels  projected on  the pupil are  much smaller  than the
Fresnel  radius, the  number of  photons per  pixel received  from even
bright stars in a short 1-ms  exposure time would be small compared to
the detector readout noise.  The EM  CCDs have a readout noise of less
than  one electron  (el)  and allow  noiseless  on-chip binning,  thus
alleviating the  de-magnification challenge;  this option was  used in
FASS \citep{FASS}.  However, noiseless binning is not possible in CMOS
cameras, while  their readout  noise is about  1\,el.  Hence,  a large
$k_{\rm  magn}$ is  indeed  necessary with  a  CMOS.  Maximum  optical
de-magnification is limited by the  Lagrange invariant (the product of
surface and solid angle is constant). Even with a large solid angle at
the detector, the maximum angle on the sky (i.e.  the field of view) is on the
order of  an arc-minute. Therefore, a  solid-state scintillation sensor
requires an additional guiding camera to keep the star centred in the
narrow field.

Combination  of MASS  and DIMM  in one  instrument solves  the guiding
problem (which  is done by DIMM),  at the cost of  having two parallel
systems with  separate detectors, acquisition channels,  and software.
DIMM measures  the total seeing  including the ground layer,  to which
MASS is insensitive.  This combination is a consequence of the physics
of optical  propagation.  Phase distortions of  optical waves produced
by atmospheric fluctuations of the  air refractive index are partially
converted  into  amplitude  fluctuations  (scintillation)  only  after
propagation.    Therefore,  any   scintillation-based  instrument   is
intrinsically  insensitive to  turbulence  near the  ground, and  only
instruments sensitive to phase distortions (like DIMM) can measure the
total seeing.

In reality,  this distinction  between a  phase-sensitive DIMM  and an
amplitude-sensitive MASS is  blurred because DIMM is  also affected by
amplitude fluctuations to  some extent and its  measurements depend on
the propagation  distance \citep{Tok2007,Kornilov2019}.  On  the other
hand, by measuring scintillation in a plane optically conjugated below
the   pupil,  a   virtual  propagation   path  is   added,  making   a
scintillation-based   instrument   sensitive    to   the   near-ground
turbulence.    This  is   the  principle   of  a   generalized  SCIDAR
\citep{Klueckers1998,Osborn2018}.     Strictly    speaking,    virtual
propagation works for an aperture of infinite size because diffraction
on  its  edges  intervenes  at  a   spatial  scale  of  the  order  of
$\sqrt{\lambda H}$  for a virtual  propagation distance $H$.   This is
not a  problem in a SCIDAR  because in a large-aperture  telescope the
area  near the  edges  affected by  diffraction  is relatively  small.
However,  in  a small  telescope  the  diffraction affects  the  whole
aperture  and  must  be  taken  into  account  explicitly  if  virtual
propagation is used.  Moreover, a defocused pupil image is affected by
the telescope shake, aberrations, and low-altitude (local) turbulence,
whereas  a pupil-based  scintillation sensor  is immune  to all  these
effects.

Extending the scintillation-based technique to measure the full seeing
is  an attractive  choice that  allows to  get rid  of the  DIMM. This
option  is currently  explored by  the FASS  team \citep{FASS2}.   The
pupil  image conjugated  to  some  distance $H$  below  the ground  is
obtained  simply by  defocusing the  telescope. When  the detector  is
placed  at a  distance $\Delta$  in front  of the  focal plane,  it is
conjugated  to $H  = F^2/\Delta$,  where  $F$ is  the telescope  focal
distance.  By using a  moderate-sized telescope of $D=30$\,cm diameter
and conjugating  to $H$  of a  few hundred metres,  it is  possible to
select an  annular zone in  the defocused  image that is  not strongly
distorted by diffraction and thus  is suitable for Fourier analysis of
intensity  fluctuations   in  the   angular  coordinate,  as   in  the
aperture-conjugated FASS.

On the other hand, when the two apertures with prisms that produce two
images in a classical DIMM are replaced by a weak conic lens (axicon),
a ring-like image is formed at the focal plane. Such extension of DIMM
to  the full  aperture increases  its sensitivity  and eliminates  the
intrinsic asymmetry of a two-aperture DIMM.  The radius of the ring is
a measure  of the defocus, and  its fluctuations caused by  turbulence are
directly  linked  to  the  seeing.  Moreover,  the  speed  of  defocus
variation allows  us to measure  the atmospheric time constant  -- an
important parameter  that affects  AO systems and  interferometers.  A
turbulence monitor based on ring images was called FADE (FAst DEfocus)
and tested using a 35-cm telescope \citep{FADE}.

The proposed  concept is a fusion  of previous ideas.  It  is based on
fast registration  of ring-like images,  as in FADE.   Fluctuations of
intensity along the ring (in the angular coordinate) reflect intensity
variation in the  annular aperture and are analogous in  this sense to
FASS. On the  other hand, the ring is focused  in the radial direction,
and radial motion of its segments gives a measure of the total seeing,
as in DIMM.  A descriptive  acronym RINGSS (Ring-Image Next Generation
Scintillation  Sensor) is  chosen for  this concept  to underline  its
descendence from MASS and FASS.

Propagation of optical waves through turbulent atmosphere is a complex
phenomenon. Moreover, the standard description of turbulence refers to
its idealized  statistical model.  As a  result, turbulence parameters
cannot  be  defined  or  measured with  arbitrary  accuracy,  and  all
instruments measuring optical turbulence  are based on approximations.
This  point, further  elaborated in  \citet{Tok2007}, is  important to
bear in mind. That paper,  among many others, lists  the definitions
of standard turbulence parameters -- the seeing $\epsilon_0$, the
Fried radius $r_0 = 0.98 \lambda/ \epsilon_0$, the refractive-index structure
constant $C_n^2$, etc. 

The    operational   principle    of    RINGSS    is   presented    in
Section~\ref{sec:principle}.  Its  main component -- the  algorithm to
measure  turbulence  parameters from  ring  images  -- is  exposed  in
Sections~\ref{sec:proc} to \ref{sec:turb}. On-sky tests of a prototype
instrument are  covered in  Section~\ref{sec:proto} and the  choice of
instrument   parameters    in   Section~\ref{sec:instr}.    Discussion
 of this new method  in Section~\ref{sec:disc}  closes the paper.

\begin{figure}
\centerline{
  \includegraphics[width=8.5cm]{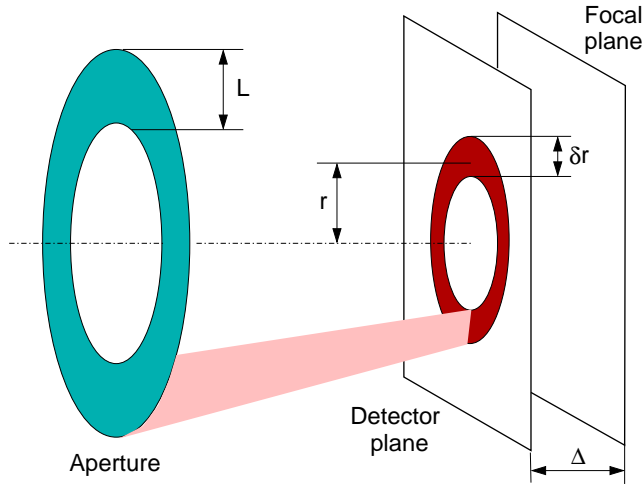}
}
\caption{Principle of the ring-image  turbulence monitor. Light from a
  single  bright star  is collected  by  a telescope  with an  annular
  aperture and some conic aberration.  The image is a ring with 
    angular radius  $r$ and  angular width  of $\delta r =
  \lambda/L$ set  by diffraction, where $  L = D (1  - \epsilon)/2$ is
  the width of the  aperture annulus.  Fluctuations of intensity along
  the  ring  are mostly  produced  by  amplitude  fluctuations at  the
  aperture (scintillation),  while deformation  of the ring  is caused
  mostly  by  phase  aberrations.   Statistics of  these  fluctuations
  enable measurement of the turbulence profile and seeing.
  \label{fig:ring} }
\end{figure}

\section{Operational principle of RINGSS}
\label{sec:principle}

The method  of signal analysis  developed here is applicable  to three
kinds  of data:  (i) images  of annular  telescope pupil,  (ii) simple
defocused images,  and (iii) defocused images sharpened  in the radial
direction.   The last  option appears  to be  the best  choice  and is
considered here, while the pupil-plane  case (i) is analogous to FASS.
The    geometry    and   basic    parameters    are   introduced    in
Fig.~\ref{fig:ring}.  An  annular  aperture  of diameter  $D$  with  a
central   obscuration  $\epsilon$   has   a  width   of   $L=  D(1   -
\epsilon)/2$.  The angular width  of a  focused ring  image is  set by
diffraction  to $\delta r  \approx \lambda/L$  (at $\epsilon  > 0.5$).
The   ring   angular  radius   is   $r_{\rm   ring, rad}   \approx  D(1   +
\epsilon)/(4H)$, where  $H=F^2/\Delta$ is the  conjugation distance of
the detector ($H=\infty$  at the focal plane) and  $F$ is the efective
focal length. By selecting a larger defocus $\Delta$, we get a smaller
$H$ and a larger ring, with less photons per pixel.

\begin{figure}
\centerline{\includegraphics[width=8.5cm]{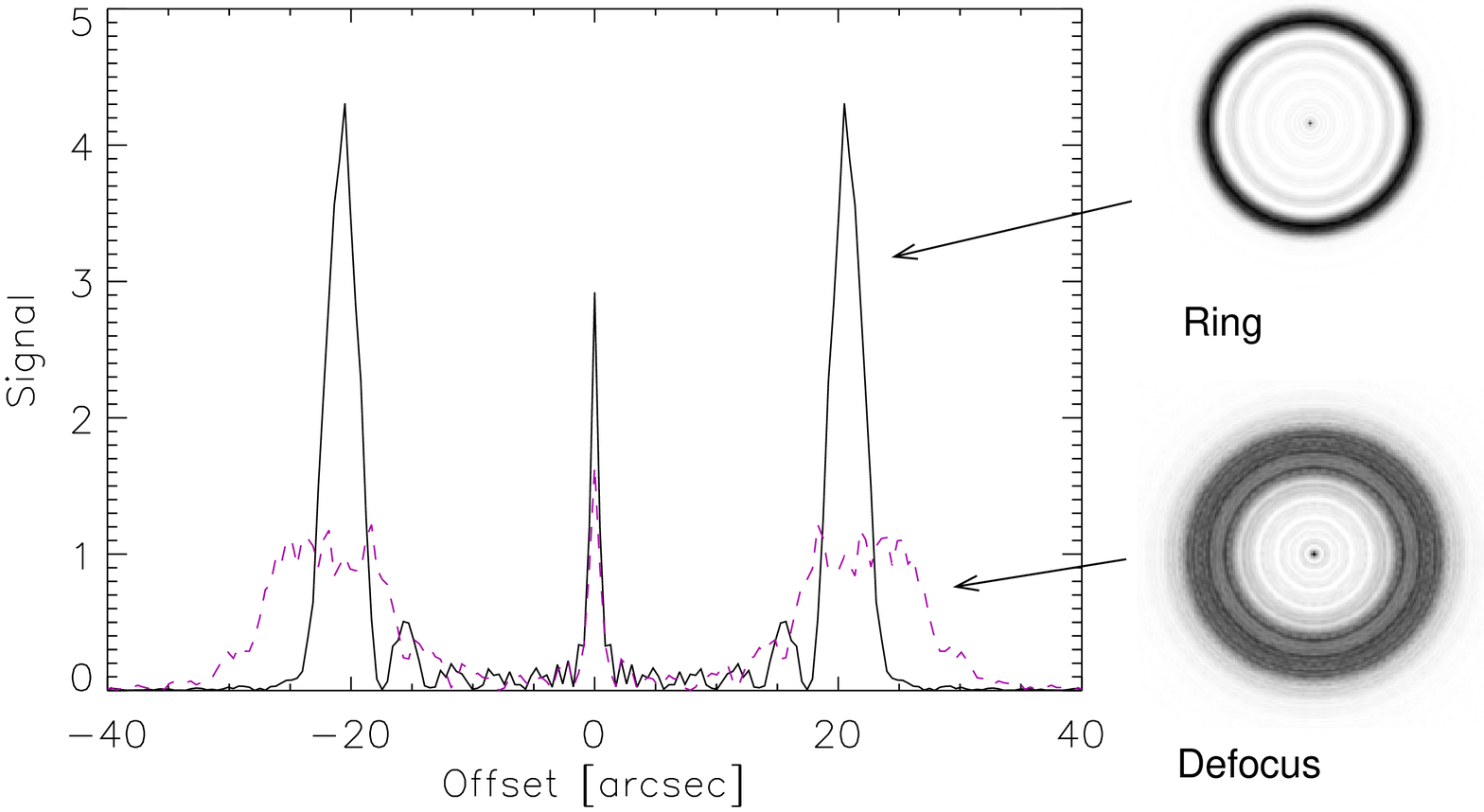} }
\caption{Comparison  between  the focused  ring  produced  by a  conic
  wavefront and  an equivalent defocused  image. The curves  show cuts
  through the images ($D=0.13$\,m, $\epsilon =0.5$, $H = 0.4$\,km).
\label{fig:defocus} }
\end{figure}

A  defocused image  in  a  small telescope  with  annular aperture  is
heavily   affected  by   diffraction   and,  therefore,   is  a   poor
representation of the pupil image. When the wavefront is conic, rather
than  spherical, the  ring  is  focused in  the  radial direction  and
becomes sharper.  Figure~\ref{fig:defocus}  compares these situations.
The maximum  signal in  the radially  focused ring  is $\sim$3$\times$
larger than in  the similarly defocused image. Only  when the aperture
is a narrow  annulus (i.e.  with a large  $\epsilon$), the distinction
between  conic and  spherical wavefronts  within the  aperture becomes
negligibly small.  A practical way to obtain a nearly conic wavefront is
to  combine defocus  with  a small  spherical  aberration of  opposite
sign.  The  spherical  aberration  compensates the  curvature  of  the
defocused  wavefront within  annular  aperture and  renders it  almost
conical.   As shown  by  \citet{FADE}, an  excellent approximation  is
achieved when the ratio of rms spherical to rms defocus aberrations is
1:10.  Such  condition can  occur when  a positive lens  is used  as a
focal reducer.

The linear  size of  the detector  pixel $p$ and  the need  to collect
enough photons per pixel favor a  short effective focal length $F$.  A
defocused  image approximates the pupil  with  a de-magnification  factor
$k_{\rm magn}  \approx H/F$.  Hence, for  a given $H$, a  small $F$ is
needed to get a large $k_{\rm magn}$.  This can be achieved by a focal
reducer  --  an achromatic  positive  lens  placed  in front  of  the
detector.  The lens also produces spherical aberration of correct sign
to make a conic  wavefront. On the other hand, the  need to sample the
ring in  the radial  direction by  at least  two pixels  restricts the
minimum  focal  length to  $F/D  >  (p/\lambda)  (1 -  \epsilon)$,  or
$F/D>2.4$   for  $p=2.9$   $\mu$m,   $\lambda  =   0.6$  $\mu$m,   and
$\epsilon=0.5$.

The  RINGSS  concept is  generic  and  independent of  the  instrument
parameters chosen to  illustrate it.  Most numerical  examples in this
paper refer to  a telescope of 0.13\,m diameter with  $\epsilon = 0.5$
and a detector conjugated to $H=400$\,m, resembling the actual prototype
described below.  Optimum sampling of  the $r_{\rm ring} \approx 25''$
radius ring at $\lambda = 0.6$\,$\mu$m calls for the relatively coarse
pixels of $1.9''$, hence a radius of $\sim$13 pixels.

A ring image is used to compute  several {\em signals} $a_m$ as sums of
the products  of  the  pixel  values   $I_i$  and the {\em  masks}  $M_{m,i}$,
normalized by the total flux $I_0$:
\begin{equation}
a_m =  I_0^{-1} \sum_{i} M_{m,i} I_{i} .
\label{eq:a}
\end{equation}
This formulation is very general. It applies to a DIMM instrument
where the signals $a$ are spot centroids  and the masks are defined
accordingly. A Fourier transform of the image can also be viewed as
a particular case of (\ref{eq:a}) where $M$ is a complex exponent
and the signals $a$ also are complex numbers.  

Use  of  the Fourier  transform  for  statistical analysis  of  random
intensity distributions appears to be  a natural choice.  However, the
scintillation pattern  in  a  small  telescope  is  truncated  by  its
aperture, and the spatial power spectrum of the intensity distribution
at  the pupil  is a  heavily aliased  version of  the intrinsic  power
spectrum of  the infinite  scintillation pattern.  Fourier  analysis of
scintillation  in an  annular  aperture   avoids  aliasing   when  polar
coordinates $(r,\theta)$ are used.  This idea, first implemented in
FASS \citep{FASS},  is exploited here. The Fourier transform of a ring
image in the angular coordinate is  a set of complex signals $a_m$ for
angular frequencies $m=0,1,2, \ldots$  that correspond to (\ref{eq:a})
with masks
\begin{equation}
M_m(r, \theta) = f_r(r) {\rm e}^{{\rm i} m\theta} .
\label{eq:ang}
\end{equation}
Here, $f_r(r)$ is the radial weight needed to reduce the impact of noise
in image pixels outside the ring  image or pupil.  The maximum angular
frequency $m$ is  set by the Nyquist limit, i.e.   at least two pixels
along the  ring per period.  This leads to  $m_{\rm max} =  \pi r_{\rm
  ring}$, with  the ring  radius in pixels.   A typical  value $r_{\rm
  ring} =10$ pixels corresponds to $m_{\rm max} \approx 30$.

The mean  square modulus  of the angular signals computed  for a
series of images  is called {\em angular power  spectrum} (APS) $S_m$,
in direct analogy to the standard power spectrum:
\begin{equation}
S_m = \langle | a_m |^2 \rangle = \sum_{j=1}^N W_m(z_j) J_j .
\label{eq:wf}
\end{equation}
The right-hand  side of (\ref{eq:wf})  relates the APS to  the optical
turbulence profile (OTP), represented  by a collection of $N$ discrete
layers at distances $z_j$ from the instrument with turbulence strength
$J_j = C_n^2(z_j) {\rm d}  z$ in each layer.  Turbulence integrals $J$
are measured in  m$^{1/3}$; a seeing of $1''$  (at 500\,nm wavelength)
corresponds  to  $J=6.83   \times  10^{-13}$\,m$^{1/3}$.   The  seeing
$\epsilon_0$  and  the  Fried   radius  $r_0$  are  common  turbulence
parameters   uniquely  related   to   $J$  for   a  given   wavelength
\citep[e.g.][]{Tok2007}.   An infinite turbulence  outer scale is
  assumed throughout  this paper because  at the small  spatial scales
  relevant here the outer scale effects are negiligible.

Equation  (\ref{eq:wf}) assumes  that each  turbulent layer  gives its
independent  contribution to  the APS,  proportional to  $J_j$ with  a
coefficient  $W_m(z_j)$ called  {\em  weighting  function} (WF).  This
assumption  holds  in the  weak  scintillation  regime. Modelling  the
combined  effect of  turbulence  by a  linear  combination of  effects
produced by all  layers is the cornerstone principle  of all turbulence
monitors, without exception.  In the case of a DIMM,  for example, the
WFs are often assumed to be independent  of $z$ and the DIMM signal is
related to the total turbulence integral, i.e. to the seeing.
 
\begin{figure}
\centerline{\includegraphics[width=8.5cm]{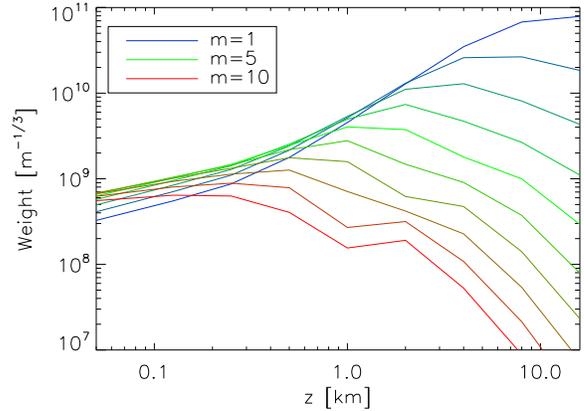} }
\caption{Weighting   functions    for   the   APS   from    $m=1$   to
  $m=10$. Instrument parameters: $D=0.13$\,m,  $\epsilon =0.5$, $H
  = 400 $\,m,  polychromatic  light (a star  of  effective  temperature
  7500\,K).
\label{fig:weightplot} }
\end{figure}

The APS  at $m=0$ is the  normalized variance of the  total flux, also
called scintillation index.   With increasing $m$, the  APS isolates a
band  of   increasing  spatial  frequencies  from   the  scintillation
spectrum. As the  characteristic scale of the scintillation  is of the
order of  $\sqrt{\lambda z}$,  a small-scale scintillation  (i.e.  large
$m$) is produced mostly at  small $z$, while a large-scale scintillation
corresponds   to   a  large   $z$.    A   set   of  WFs   plotted   in
Fig.~\ref{fig:weightplot} illustrates their dependence on $z$ and $m$:
$S_{10}$ is mostly  sensitive to turbulence within  1\,km, while $S_1$
measures  mostly  the  distant  turbulence.  Owing  to  the  different
$z$-dependence  of the  WFs, equation  (\ref{eq:wf}) can  be inverted,
solving  for  a set  of  turbulence  integrals  $J_j$ that  match  the
measured APS $S_m$. This is  the common operational principle of MASS,
FASS,  and RINGSS.   The differences  between  them are  in the  input
signals and in the WFs.

In addition to  measuring the OTP from  scintillation, RINGSS provides
an  alternative estimate  of the  seeing from  the differential  image
motion, as  in a DIMM.  This  extra benefit is not  available when the
telescope is simply defocused  without radial ring sharpening.  Radial
deformations  of  the ring  are  estimated  similarly to  the  angular
coefficients, i.e.   by summing  products of  pixel values  and masks.
The ring  is divided into eight  $45^\circ$ sectors and the  radius in
each  sector is  computed by  the centroid  algorithm a  with suitable
radial  mask (eq.~\ref{eq:rk}  below).   The  rationale for  computing
radii by sectors is twofold.   First, it gives four DIMM-like signals
of  longitudinal  distances between  opposite  sectors  (sum of  their
radii)  for measuring  the  seeing.  Second,  it  allows to  determine
the position  of the  ring  centre from  the  difference between the opposite
radii.  This provides a robust way to centre the ring in each frame.

Summarizing, the operational principle of RINGSS is similar to that of
MASS and  FASS. Series of  ring images  are recorded and  processed to
extract the angular signals $a_m$  and the sector radii. The variances
of these quantities are interpreted in terms of the turbulence profile
$J_j$  by  solving  a  system of  linear  equations  with  appropriate
WFs. The two key components of RINGSS are the algorithms for computing
the signals and  the calculation  of the WFs; they are covered
in the following  Sections. These algorithms were  verified using both
simulated and real data.

\section{Image processing}
\label{sec:proc}

\begin{figure}
\centerline{
  \includegraphics[width=8.5cm]{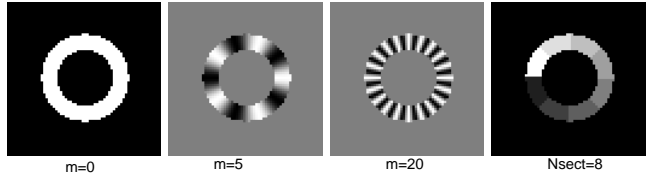}
}
\caption{Spatial  masks  applied  to   the  64$\times$64  pixels  ring
  images. The first  panels show three cosine filters,  the last panel
  is a sum of the sector  masks $F_k$, each multiplied by $k$ to distinguish
  the sectors.
\label{fig:masks} 
}
\end{figure}

Series  of consecutive  ring  images (image  cubes)  are processed  to
determine  angular signals (or coefficients) $a_m$,   sector radii $r_k$, and
several additional  parameters   such  as   background,  ring   radius,  noise
estimates,  etc. The  image cubes  can be  recorded on  a disk  by the
acquisition  software or  passed to  the processing  pipeline directly
without saving, as done typically in a DIMM. Series of  signals
extracted from the image cubes are then used to compute statistical moments
(variances and covariances) needed for the OTP measurement. 

\subsection{Calculation of the signals}

Spatial  masks  have   the  same  size  as  the   image  frames,  e.g.
64$\times$64 pixels.  The  masks for computing the  real and imaginary
(cosine and sine) parts of the angular signals $a_m$ are given by
eq.~\ref{eq:ang}. The  radial mask  $f_r(r)$ equals  one for  all pixels
with radii within the mask half-width  $\Delta r$ from the ring radius
$  r_{\rm ring}$  (in  pixels)  and zero  otherwise.   A smooth  (e.g.
Gaussian) $f_r(r)$ was tested, but  the simple sharp radial cutoff works
quite well.   The default choice  is $\Delta  r= 1.5 \delta  r$, where
$\delta  r$ is  the full  width at  half maximum  (FWHM) of  the ring,
i.e. its thickness.   Using a smaller mask width  decreases the noise,
but  produces  biased  results   (e.g.   under-estimated  seeing),  as
demonstrated by  simulations, while a  wider mask increases  the noise
without    affecting    other   parameters.     Figure~\ref{fig:masks}
illustrates the cosine masks with $m=0, 5, 20$.

For  measuring the  ring radius,  we divide  it into  $N_{\rm sect}=8$
radial sectors, each sector covering $45^\circ$ in angle. For a sector
$k$, the  mask $F_k = f_r(r)$  within the sector and  zero otherwise; it
measures the total signal (flux) within the sector.  The mask $R_k = r
F_k$ estimates the ring radius in this sector:
\begin{equation}
r_k = \sum_{i} R_{k,i} I_{i}  / \sum_{i} F_{k,i} I_{i} .
\label{eq:rk}
\end{equation}

\subsection{Centring the rings}

Each  frame must  be centred  and  the background  must be  subtracted
before computing  the products of  pixel values and masks.   Also, the
ring  radius  and width  must  be  known  to  define the  masks.   The
parameters $  r_{\rm ring}$ and  $\delta r$ can  be fixed for  a given
instrument, but currently  they are estimated `on the  fly' from the
average first 50  frames of the image cube, to  reduce the noise. The
background  is also  estimated from  the average  image as  the median
value of all its pixels outside $ 1.5 r_{\rm ring}$.

Centring  of each  frame is  a critical  part of  the algorithm.  The
sector radii  computed by (\ref{eq:rk})  give an estimate  of the residual
ring offsets $dx$ (similarly $dy$) as
\begin{equation}
  dx = (2.3/N_{\rm sect}) \sum_k r_k \cos \theta_k, 
\label{eq:xc}
\end{equation}
where  $ \theta_k$  are  the  angles of  the  sector  centres, and  the
proportionality   coefficient  2.3   was   determined  by   processing
artificially shifted  images. The offsets determined  from the current
frame are applied to centre the following frame, so accurate tracing
of the ring  motion is achieved. This algorithm was found to be
very robust.

The frames  should be  centred within  a fraction of  a pixel.   When the
centring  is done  only within one  pixel (by  integer image  shifts), the
estimated  centres are  biased to  integer  values, so  the curves  of
displacement vs.   frame number computed for simulated  image cubes  with ring
motion are distorted.  Moreover,  integer shifts introduce a sub-pixel
jitter of the  otherwise static ring and cause  fluctuations of the
differential  sector radii  and the angular  signals that  exceed the
photon-noise errors  at high flux.  Fractional sub-pixel shifts  by bilinear
interpolation  were tested  to fix  this issue. However,  interpolation
introduces correlation  between pixels  and affects  the shape  of the
noise spectrum,  damping it  at high frequencies.   Finally, sub-pixel
shifts by  Fourier transform  are implemented.   Unlike interpolation,
they do not distort the noise statistics.

\begin{figure}
\centerline{
\includegraphics[width=8.5cm]{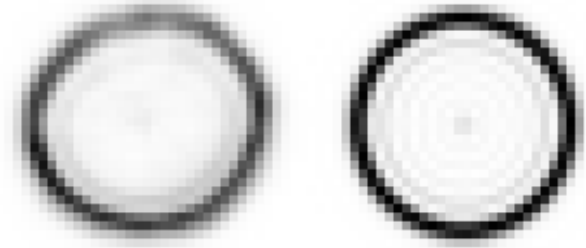} }
\caption{Comparison  of  the  average   centred  ring  image  in  the
  prototype instrument (left) and the simulated undistorted image with
  matched parameters (right).
  \label{fig:imagecompare} 
}
\end{figure}

Re-centred  frames are  summed up  (averaged) and  saved for  off-line
examination.    Figure~\ref{fig:imagecompare}  compares   the  average
re-centred image  from a  real instrument  with the  ideal diffraction
image.  Note  the similarity  of the  diffraction structure,  e.g. the
faint inner  ring and a small  spike at the centre.   The real average
ring is wider, being affected by distortions under a $2''$ seeing. Its
non-uniformity  in  azimuth  is   caused  by  small  residual  optical
aberrations of the telescope, in particular coma and astigmatism.

To compute the radii for $N_{\rm sect}=8$ sectors and 31 $a_m$ complex
signals for  $0 < m <  m_{\rm max}=30$, a  total $2 N_{\rm sect}  + 2(
m_{\rm  max}  +1)  =  78$   masks  are  needed.   Calculation  of  all
signals is implemented as  a matrix-vector product.  Pixel values
$I_i$  in   a  64$\times$64  frame  (after   centring  and  background
subtraction) are  arranged in  a single  4096-element vector,  and all
masks are  combined in a  78$\times$4096 mask matrix. However,  with a
sharp radial  mask, only  the pixels  inside the  mask are  used.  This
reduces  the dimension  of the  mask  matrix from  4096 to  $\sim$1000
elements  per  line  and  speeds up  the  calculation.   For  example,
processing a cube of 64$\times$64$\times$2000 format on a laptop takes
1.06\,s with the full matrix and only 0.44\,s with the reduced matrix.

Processing of the  image cube results in a matrix  of the signals
and returns additional  parameters such as the ring radius  and width, its
average centre  (useful for automatic guiding),  background, flux, and
the factors $\nu$ needed to estimate the noise (see below).

\subsection{Noise estimation}
\label{sec:noise}

Without turbulence, the ring image is static (it only moves as a whole
owing  to  the  telescope  tracking errors),  but  the  signals  still
fluctuate because of the photon  and readout noise. Comparison between
analytic estimates of the noise  variance and the variance computed on
simulated cubes is  a good test of the algorithm.   The noise variance
(bias) must be subtracted from the  measured APS and from the variance
of  the  differential sector  motion  for  correct estimation  of  the
atmospheric parameters.

Let $I_i$  be the pixel  values in photo-electrons. Their  variance is
$I_i +  R^2$, where  $R$ is  the readout  noise in  electrons, assumed
to be equal  in all  pixels.  To  estimate   the APS of the noise  (i.e. the
variance   of  the   angular   signals),  the   mask  values   in
eq.~\ref{eq:a}  must  be  divided  by  the  normalization  factor
$N_{\rm ph} =  \langle \sum_i f_r(r) I_i  \rangle$, i.e. the  photon flux
within the radial  weight $f_r(r)$,  typically a 0.9 fraction of  the  full
flux. Then,  taking advantage of  the uncorrelated pixel noise,  we sum
their contributions to the variance and express the result as
\begin{equation}
S_{\rm noise}  = \langle |a|^2_{\rm  noise} \rangle = \nu_1/N_{\rm  ph} +
\nu_2 (R/N_{\rm ph})^2,
\label{eq:anoise}
\end{equation}
where the  factors $\nu_1$ and $\nu_2$  characterize the contributions
of  the photon  and readout  noise, respectively.  They depend  on the
radial mask $f_r(r)$ and  on the ring shape, but do  not depend on the
flux and readout noise:
\begin{eqnarray}
  \nu_1 & = &  (1/N_{\rm ph}) \; \sum_i f_r^2(r_i) I_i,  \nonumber \\
  \nu_2 & = &  \sum_i f_r^2(r_i) .
  \label{eq:nu}
\end{eqnarray}  

Here the fact  that $\langle |a|_{\rm noise}^2 \rangle$ is  the sum of
the cosine and  sine variances is used, eliminating  the dependence on
$m$:  the  theoretical noise  spectrum  is  flat.   The sense  of  the
noise factors $\nu$ is very clear when $f_r(r)$ takes only values of 1 or
0. Then $\nu_1  =1$ and $\nu_2$ is the number  of pixels with non-zero
weight, $N_{\rm  pix}$.  The  noise variance  is then  $(1 +  R N_{\rm
  pix}/N_{\rm ph})/N_{\rm  ph}$.  The second  term in the  brackets is
the ratio  of the readout noise  to the average number  of photons per
pixel.  When this  ratio is $\ll$1 (which is the  typical situation in
practice), the photon noise dominates,  otherwise the readout noise is
the main contributor to the noise variance of the signals.

The noise  variance of the  differential sector motion (in  pixels) is
computed in a  similar way by replacing  in Eq.~\ref{eq:nu} quantities
$f_r^2(r_i)$ with  $f_r^2(r_i) (r_i  - r_{\rm ring})^2$.   The factor
$\nu_{1r}$  expresses the  variance due  to  the photon  noise and  is
approximately equal to  the square of the ring radial  rms width (FWHM
divided by 2.35) in pixels.  The factor $\nu_{2r}$ is larger than
$\nu_{2}$  because  the  calculation   of  radius  weights  pixels  in
proportion to  $(r -  r_{\rm ring})^2$  and effectively  increases the
relative importance of the readout  noise. For example, in a simulated
data with a sharp radial mask of  the width $\Delta r = 1.5 \delta r$,
the noise  factors $\nu_1,  \nu_2, \nu_{1r}, \nu_{2r}$  are $1.0,
631, 1.02, 1952$,  respectively.  The four noise  factors are computed
during the image cube processing using the average re-centred image to
estimate $I_i$.

Static ring images  distorted only by the noise  (with optional motion
of  the ring  as a  whole over  the detector)  were simulated  and the
resulting data cubes were processed  by the algorithm described above.
Variances of the angular signals and of the differential sector motion
were found  to match  the noise  estimates very  well, within  10 per cent or
better.   Furthermore, noise  estimates  were  checked using  physical
simulation of  a static ring  image projected  on to the  camera.  The
number  of photons  varied by  varying both  the illumination  and the
exposure time.  Again, a good  agreement between the estimates and the
actual measurements was found.

It  is instructive  to compare  the noise  variance with  the expected
scintillation  variance.  The  $m=1$ WFs  in Fig.~\ref{fig:weightplot}
increases  by  two orders  of  magnitude  with increasing  propagation
distance, but the $m=10$ WF  peaks at $\sim 5 \times 10^8$\,m$^{-1/3}$
at small distances  and decreases further out. A  weak turbulent layer
at  the ground  with  $J =  2.1  \times 10^{-13}$\,m$^{1/3}$  ($0.5''$
seeing) would produce an APS  of $S_{10} \sim 10^{-4}$.  It equals the
photon  noise  for  $N_{\rm  ph}  =  10^4$  which,  in  our  prototype
instrument, is a flux  from a $V \sim 2$ mag star  in a 1-ms exposure.
Therefore,  subtraction of  the  noise bias  is  critical for  correct
measurement of  the weak turbulence near the  ground by scintillation,
whereas  the  signal at  small  $m$,  dominated  by the  high-altitude
turbulence, always largely exceeds the noise. Anyway, the photon noise
in  RINGSS is much  smaller than  in MASS,  where its  calculation and
subtraction is even more critical.

\section{Weight calculation}
\label{sec:wf}

Calculation of the  theoretical WFs needed to interpret  the signal of
RINGSS is a vital part of the proposed technique. A WF of any turbulence
monitor is  computed by  the standard  method as  the integral  of the
product of the turbulence  phase power spectrum $\Phi_\varphi(f)$, the
propagation filter $P(f)$, and the instrument frequency filter $Q({\bfit
  f})$:

\begin{equation}
W = \int {\rm d}^2 {\bfit f} \; \Phi_\varphi(f) P(f) Q({\bfit f}) ,
\label{eq:W}
\end{equation}
where ${\bfit  f}$ is the  two-dimensional spatial frequency and  $f$ is
its modulus.  The phase power  spectrum is radially symmetric and, for
Kolmogorov   turbulence,   is   given    by   the   standard   formula
\citep{Tatarskii,Roddier81}
\begin{equation}
\Phi_\varphi(f) = 0.00969 (2 \pi/\lambda)^2 f^{-11/3} J ,
\label{eq:Phi}
\end{equation}
where the turbulence integral $J= C_n^2 {\rm d}z$ should be set to one
in order to  compute  the  WF.  The propagation  filter  for  scintillation  in
monochromatic light,
\begin{equation}
P(f) = \sin^2 (\pi \lambda z f^2),
\label{eq:P}
\end{equation}
is also radially symmetric.  Its generalization to polychromatic light
is given by \citet{poly}.  In the  case of MASS, the instrument filter
is  the  square modulus  of  the  Fourier  transform of  the  aperture
function,  normalized   to  the  unit  area.\footnote{In  a DIMM,  the
instrument filter  is not axially symmetric,  while the propagation
is often neglected \citep{Kornilov2019}.}

\begin{figure}
\centerline{\includegraphics[width=8.5cm]{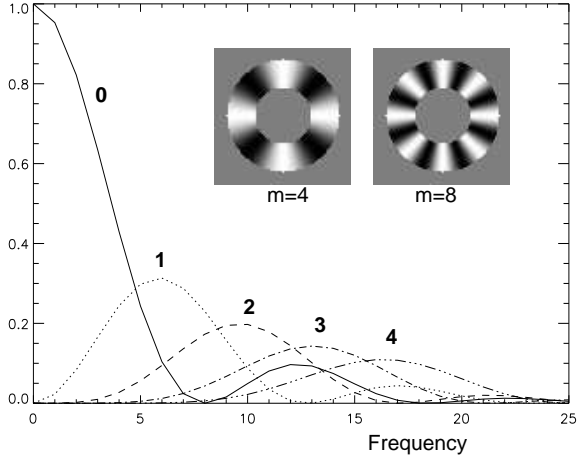} }
\caption{Frequency  filters $Q_m(f)$  for  the  annular aperture  with
  $\epsilon = 0.5$.  The inserts show the cosine filters for $m=4$ and
  $m=8$.  All  filters  are  normalized  to  the  same  integral.  The
  frequency axis is in arbitrary units.
\label{fig:resp} }
\end{figure}

When  the scintillation  is recorded  at the  pupil, as  in FASS,  the
function  $f_r(r)$   in  eq.~\ref{eq:a}  corresponds  to   the  aperture
transmission, and the instrument response  filter for the APS $S_m$ has
an analytic expression
\begin{equation}
Q_m(f) = \left( 2 \pi \int_0^\infty f_r(r) J_m(2 \pi f r) r {\rm d}r
\right) ^2 ,
\label{eq:Hankel}
\end{equation}
which  is the  square  of the  $m$-th order  Hankel  transform of  the
aperture   function    $f_r(r)$.    Figure~\ref{fig:resp}    shows   how
aperture-plane filters  with increasing $m$ isolate  different spatial
frequency  bands.  These  signals  are analogous  to the  differential
scintillation in MASS that also senses a certain frequency band of the
scintillation  power   spectrum,  cutting   out  both  low   and  high
spatial frequencies.

The ring images are sensitive to both amplitude and phase fluctuations
at  the   pupil.   At  large  propagation   distances,  the  amplitude
fluctuations dominate,  and the WFs of  RINGSS and aperture-conjugated
FASS are  very similar, as  shown below.  However, at  small distances
the  phase  fluctuations cannot  be  neglected.   Phase and  amplitude
fluctuations at the  pupil are mutually correlated and  are subject to
different instrumental frequency  filters. It is thus  not possible to
treat phase and amplitude separately; instead, their joint effect
on  the signal  variance  should be  evaluated.   The propagation  and
instrument filters are interwined, so the combined response filter
$PQ$ in eq.~\ref{eq:W} is computed.

\begin{figure}
\centerline{
\includegraphics[width=8.5cm]{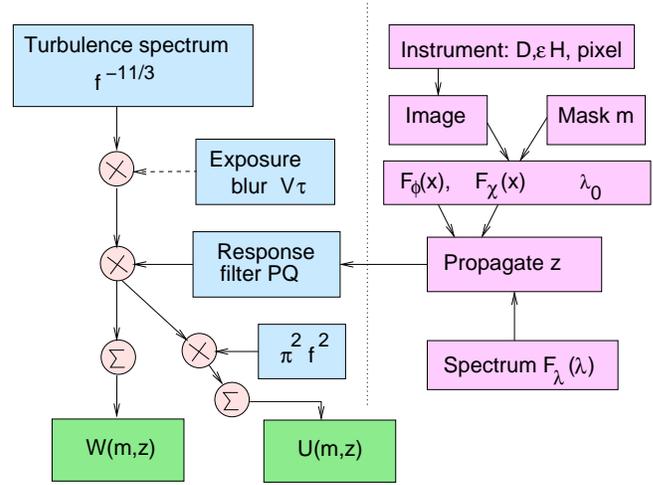} }
\caption{Flow chart of the analytic  WF calculation. Blue boxes are 2D
  arrays in the spatial  frequency domain approximating the turbulence
  spectrum  and various frequency  filters. The  WFs are  estimated as
  integrals (sums)  over the  frequency.   The  functions $U(m,z)$
    are defined in Section~\ref{sec:temp}. 
  \label{fig:wfcalc} 
}
\end{figure}

One  obvious way  to tackle  the problem  is by  numerical simulation.
However, the  WFs estimated in  this manner contain  statistical noise
and their calculation is not  fast, especially for polychromatic light
where  each wavelength  must be  treated separately;  the polychomatic
signal $a_m$ is  the average of monochromatic signals  weighted by the
spectral response $F_\lambda$.

The  analytical  formula describing  a  small-signal  response of  a
general turbulence sensor comes to  the rescue. This formula, given in
the Appendix  of \citet{Tok2007},  has been  applied  in the  past to  the
analysis  of DIMM  and FADE.   Small  fluctuations of  the signal  $a$
derived from  the product of  the image and  mask $M({\bfit x})$,  as in
eq.~\ref{eq:a}, equal
 \begin{equation}
\Delta a =   \int  {\rm d}^2
       {\bfit x} \; F_{\varphi} ({\bfit x}) \; \varphi ({\bfit x}) \;  +   
 \int  {\rm d}^2 {\bfit x} \; F_{\chi} ({\bfit x}) \; \chi ({\bfit x})  ,
\label{eq:delta1}
\end{equation}
where  ${\bfit  x}$  is  the  spatial coordinate  at  the  pupil  plane,
$\varphi({\bfit x})$ and $\chi ({\bfit  x})$ are small phase and amplitude
wavefront distortions  at the pupil,  and the functions  $ F_{\varphi}
({\bfit x})$ and $ F_{\chi} ({\bfit x})$ represent the instrument response
to phase  and amplitude,  respectively. They  equal the  imaginary and
real parts of the auxiliary quantity $A({\bfit x})$ given by
\begin{eqnarray}
A({\bfit x}) & = &  2 \; (\lambda^2 I_0)^{-1} \; E({\bfit x})
\int  {\rm d}^2
      {\bfit x'}  \; E^*({\bfit x} + {\bfit x'}) \; \tilde{M}({\bfit x'} /\lambda) 
\nonumber \\
 & = & 2 ( \lambda^2 I_0)^{-1} \; E \; [ E^* \star \tilde{M} ] ,
\label{eq:A}
\end{eqnarray}
where $E({\bfit x})$  is the unperturbed complex amplitude  of the light
waves at the  pupil and $ \tilde{M}({\bfit x'}/\lambda)$  is the Fourier
transform of the mask. A  defocused image corresponds to the quadratic
wavefront, $E({\bfit x}) = \exp ( i  c |{\bfit x}|^2)$, while for the ring
image  the  wavefront is  conic,  $E({\bfit  x}) =  \exp  (  i c'  |{\bfit
  x}|)$. Arbitrary aberrations can be  accounted for by including them
in $E$.
  
The formula (\ref{eq:A}) looks  complicated, but its implementation is
straightforward. The  functions are  represented by arrays  of 512$^2$
points  (or  1024$^2$  points  in   the  case  of  larger  apertures).
Integrals are replaced by sums,  and the convolution between $E^*$ and
$\tilde{M}$ is computed via product  of their Fourier transforms.  The
flow chart of the WF calculation is shown in Fig.~\ref{fig:wfcalc}.  For a
unit   mask,   $   \tilde{M}({\bfit   x'}/\lambda)$   is   the   Dirak's
$\delta$-function times $\lambda^2$, and the response degenerates into
$ E E^*/I_0$,  i.e.  the normalized pupil  transmission function.  The
corresponding signal  $a$ is the  normalized fluctuation of  the total
flux, and  its variance is  $S_0$.  The functions $  F_{\varphi} ({\bfit
  x})$  and $  F_{\chi} ({\bfit  x})$  are computed  separately for  the
cosine and sine masks at a  given $m$, and the corresponding frequency
filters are summed in the calculation of $PQ$.

A turbulent  layer at  the pupil contains  only phase  distortions (no
scintillation).  In this  case, the  variance of  the signal  $\langle
\Delta a^2 \rangle$ is estimated as the integral of the product of the
phase power spectrum $\Phi_\varphi$ and the spectral filter $ PQ
= | \tilde{F}_{\varphi}  |^2 $. However, after propagation,  the amplitude and
phase  fluctuations are  modified, and  they are,  generally speaking,
correlated.   We   cannot  apply  the  phase   and  amplitude  filters
separately, but have to `propagate' them back to the turbulent layer
using eq. (A10) from \citet{Tok2007}:
\begin{equation}
  PQ({\bfit f}) = 
  | \tilde{F}_{\varphi}({\bfit f}) \cos (\pi \lambda z |{\bfit f}|^2)
 -  \tilde{F}_{\chi}({\bfit f}) \sin (\pi \lambda z |{\bfit f}|^2) |^2 .
 \label{eq:fresnel}
\end{equation}
This quantity is the frequency filter applied to the phase power
spectrum. The integral of their product is the WF for the monochromatic light.

\begin{figure}
\centerline{  \includegraphics[width=8.5cm]{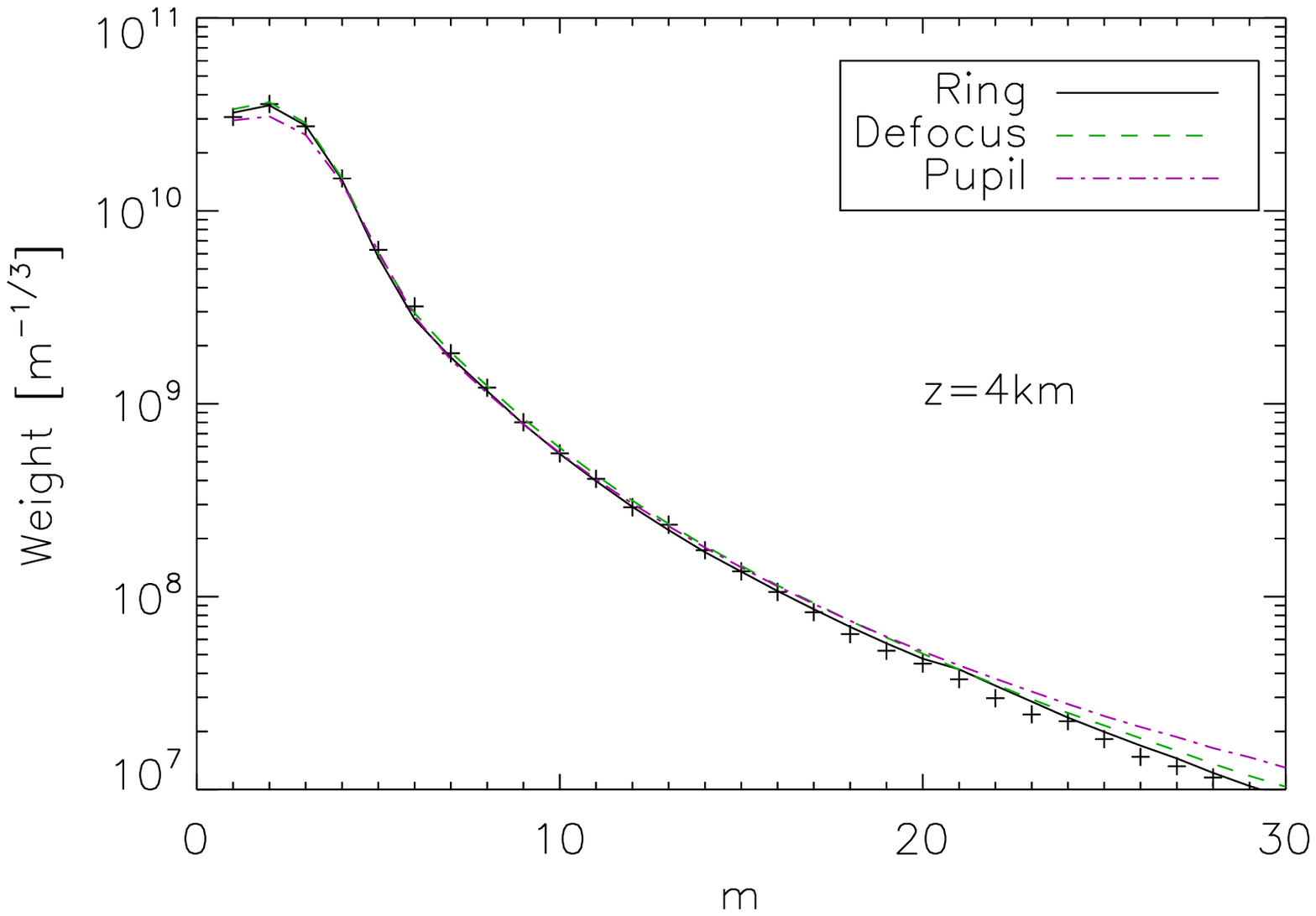} }
\centerline{  \includegraphics[width=8.5cm]{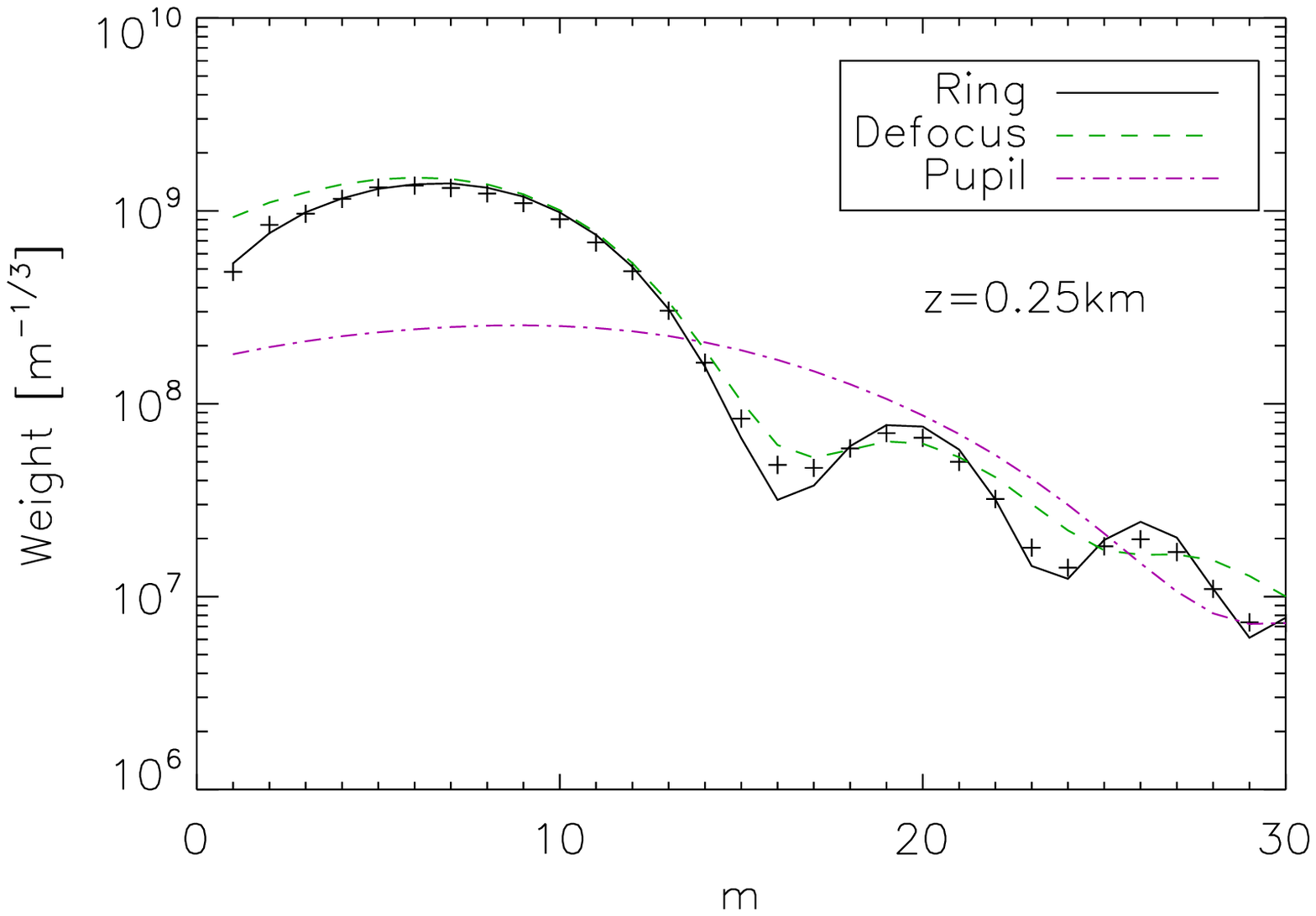}
}
\caption{Comparison  of  the  monochromatic  WFs  of  RINGSS  computed
  analytically  (solid   black  line)  and  by   numerical  simulation
  (crosses). The dashed green line is the WF for a defocused image and
  the  dash-dot   magenta  line   is  the   WF  for   the  pupil-plane
  scintillation.
\label{fig:weightcompare} 
}
\end{figure}

Figure~\ref{fig:weightcompare}  shows how  well the  monochromatic WFs
computed  analytically  match  the  results  of  numerical  simulation
(crosses)   in   the   case   of  weak   turbulence   (small-amplitude
scintillation). The  same code  works for estimation  of the WF  for a
defocused  image (suffice  to  modify the  unperturbed amplitude  $E$,
replacing the conic wavefront by the spherical one) and for the APS of
the pupil-plane  scintillation.  For a  turbulent layer at  4\,km, all
three WFs are very similar  because intensity fluctuations in the ring
or in the  defocused image resemble the intensity  distribution at the
aperture. However,  the response of an image-plane  sensor like RINGSS
to a turbulence at 250\,m is  much stronger because scintillation at the
pupil plane after such a  short propagation is very small, while phase
distortions produce a measurable effect on the ring in the image plane.

The non-monotonous dependence of the WFs  on $m$ at small $z$ (see the
lower panel  of Fig.~\ref{fig:weightcompare})  is related to  the fact
that the scintillation power spectrum  is proportional to $\sin^2 (\pi
\lambda f^2 z)$, with zeroes at  $f_n^2 = n/ (\lambda z)$.  The signal
$a_m$ isolates spatial frequencies around $ \pi D (1 + \epsilon)/(2m)$
in  the  pupil  plane   (see  Fig.~\ref{fig:resp}).   By  setting  the
propagation distance to $  z + H$, we find that the  minimum of the WF
that  corresponds to  the  first zero  of  the scintillation  spectrum
should  occur  at  $m_{\rm  min}  \approx  \pi  D  (1  +  \epsilon)/[2
\sqrt{\lambda  (z+H)}]$.   This  crude  estimate  matches  the  actual
location of the WF minima reasonably well.

In the  case of polychromatic light, the scintillation  spectrum is
damped  at  spatial  frequencies   larger  than  $1/\sqrt{\lambda  z}$
\citep{poly}.  An analytical expression in that paper was developed for a
quasi-Gaussian spectral response  $F_\lambda(\lambda)$ proportional to
$(\lambda/\lambda_0) \exp  [ -(\lambda  - \lambda_0)^2 /  2 \sigma^2]$
with $\sigma = \Delta \lambda/2.35$. In such case, the cosine and sine
terms   in   (\ref{eq:fresnel})   are  multiplied   by   the   damping
factor\footnote{ The coefficient is $1.78/2$ because the
damping factor is applied here before taking the square modulus of the
frequency filter.}  $\exp[ -(1.78/2) (\Delta \lambda)^2 f^4 z^2 ]$.

For an arbitrary spectral response, the cosine and sine terms
in  (\ref{eq:fresnel}) are replaced by the response-weighted sums, e.g.
\begin{equation}
 \cos (\pi \lambda  z |{\bfit f}|^2) \;\; \rightarrow\;\;  C \sum_k (F_{\lambda,k}/\lambda_k)
 \cos (\pi \lambda_k z |{\bfit f}|^2).
\label{eq:cos}
\end{equation}
The normalization constant $C  = 1/ \sum_k F_{\lambda,k}/\lambda_k$ is
chosen to get the unit value  of the cosine at zero spatial frequency.
This expression is accurate when the spectral response is indeed a set
of discrete  wavelengths, as  in the  simulations.  When  a continuous
response  is represented  by  a set  of discrete  values,  the sum  in
eq.~\ref{eq:cos} shows  `ringing' at  high frequencies.  The ringing is
suppressed by including the additional damping factor $\exp [-1.5 (f^2
  \delta \lambda  z)^2]$, where  $\delta \lambda$ is  the step  of the
wavelength  grid.   The  WF  calculation  for  an  arbitrary  spectral
response  was  checked  by  supplying a  quasi-Gaussian  response  and
comparing the result with the analytical formula.

The  functions $  F_{\varphi} ({\bfit  x})$  and $  F_{\chi} ({\bfit  x})$
(response to phase and amplitude)  also depend on the wavelength. This
circumstance  is  neglected here, and  they are  calculated  for  the  mean
wavelength  $\lambda_0$; only  the propagation  terms account  for the
spectral  response.  Simulations  show that  this approximation  works
well.

The WFs  plotted in  Fig.~\ref{fig:weightplot} are calculated  for the
spectral response of  the prototype instrument and  a stellar spectrum
approximated  by  a black-body  of  7500\,K  temperature.  At  angular
frequency $m=10$  and $z=4$\,km, the  polychromatic WF is  $5.3 \times
10^7$\,m$^{-1/3}$.     The   corresponding    monochromatic   WF    in
Fig.~\ref{fig:weightcompare} is $5.0  \times 10^8$\,m$^{-1/3}$, almost
an order  of magnitude larger. This  example shows that the  effect of
spectral  response is  strong and  must be  accounted for  accurately.
However, a  set of polychromatic  WFs for black-body  temperatures $T$
from 4000 to 10$^4$ K can be approximated by linear functions of $\log
T$ with an rms accuracy of 5 per cent, sufficient for practical purposes. The
influence  of the  stellar  spectrum  on the  WFs  can  be reduced  by
choosing a narrower spectral response of the instrument.

The response coefficient  of the differential sector  motion in RINGSS
is  computed  by the  same  algorithm,  only  the mask  functions  are
different.   This  coefficient  $C_r$  is normalized  to  express  the
differential sector  motion variance $\sigma^2_{2r}$ (variance  of the
sum of radii  in opposite sectors in square radians)  in one sector in
$\lambda/D$ units for $D/r_0 = 1$,
\begin{equation}
\sigma_{2r}^2  (D/\lambda)^2 = C_r (D/r_0)^{5/3} ,
\label{eq:cr}
\end{equation}
in  analogy with  the DIMM  response coefficients  \citep{PASP02}.  The
left hand side  of (\ref{eq:cr}) is a dimensionless analogue of the
APS, and the right  hand side can be viewed as a  WF by recalling that
$r_0^{-5/3}  =   0.423  (2   \pi/\lambda)^2  J$  and   setting  $J=1$.
Neglecting  for  the  moment  the  weak dependence  of  $C_r$  on  the
propagation distance, the measured  $\sigma^2_{2r}$ can be converted into
the approximate seeing $\epsilon_0$ by the formula
\begin{equation}
\epsilon_0  =   0.98    \frac{\lambda}{r_0}   \approx   0.98   \left(
\frac{\sigma^2_{2r}}{4  C_r}\right)  ^{3/5}  \left(  \frac{D}{\lambda}
\right) ^{1/5} .
\label{eq:seeing}
\end{equation}
In  fact,  the  response  of  both RINGSS  and  DIMM  depends  on  the
propagation distance  $z$.  For  the instrument considered  here, $C_r
=0.061$  at  $z=0$ and  $C_r  =0.043$  at  $z=16$\,km. The  effect  of
spectral bandwidth is negligible.

Remember that the WF calculation uses the small-signal approximation
where the phase and amplitude fluctuations are much less than one. The
validity of this regime is explored in the following Section by means
of simulation.

\section{Measurement of turbulence parameters}
\label{sec:turb}

The statistics  of the measured  coefficients $a_m$ and $r_k$  and the
knowledge of the  WFs are the two main ingredients  needed to estimate
the turbulence  parameters --  seeing, turbulence profile  $J_j$, and
the time constant. Bias introduced  by the detector noise is estimated
using the results of  Section~\ref{sec:noise} and subtracted. However,
the reduction  of variance owing to  the finite exposure time  and the
deviations from  the weak-scintillation regime should  be corrected in
order to get unbiased results.

\subsection{Temporal response of RINGSS and the atmospheric time constant}
\label{sec:temp}

The analysis of  temporal effects in this Section  closely follows the
work   by   \citet{Kornilov2011}.    He   introduced   the   quadratic
approximation  of the APS  dependence on  the integration  (exposure) time
$\tau$, valid when  the wavefront shift during the exposure,  $V \tau$, is
less than the scintillation spatial  scale ($V$ is the wavefront,
or wind, speed).  The main expression from the Kornilov's work relates
the signal variance $S(\tau)$ for a finite exposure time $\tau$ to its
zero-exposure variance $S(0)$ by the first-order quadratic formula
\begin{equation}
  S_m(\tau) \approx S_m(0) - \frac{\pi \tau^2}{6} \int C_n^2(z) V^2(z) U_m(z)
  {\rm d}z .
  \label{eq:stau}
\end{equation}
Here $C_n^2(z)$  is the turbulence  strength in m$^{-2/3}$,  $V(z)$ is
the wind speed, and $U_m(z)$ (U-functions) are analogous to the normal
WFs.   The  U-functions,  measured  in  m$^{-7/3}$,  are  computed  as
integrals  of the  product of  the turbulence  power spectrum  and the
square  modulus  of  the  spatial  filter,  but  with  the  additional
multiplicative factor of $\pi^2  f^2$ (Fig.~\ref{fig:wfcalc}).  The WF
calculation code developed  here can account for  the signal averaging
due to a linear blur of $V \tau$ metres and can optionally compute the
U-functions together with the normal WFs.

 The signal of MASS  is sampled continuously. In CMOS cameras, the
  pixels are read  sequentially during each time interval  $\tau$ by a
  `rolling shutter',  causing a pixel-dependent shift  of the sampling
  sequence.  This minor effect  is neglected  here.   The Kornilov's
theory  operates   with  the  variances  of  the   signals  and  their
covariances with a time lag of 1, assuming a regular temporal sampling
with the  cadence $\tau$.  Temporal  covariances with larger  lags are
not needed.   Let $C_{1,m} =  \langle a_{m,i} a^*_{m,i+1}  \rangle$ be
the covariance of the signal $a_m$ with a lag of one sampling interval
and $  S_m = \langle |a_m|^2  \rangle$ -- the  variance (assuming zero
mean).  Then the correlation coefficient $\rho_m = C_{1,m}/S_m$ is the
measure of  the speed of signal variation.  It is close to  1 for slow
(well-sampled) signals,  but can  be small or  even negative  when the
signal  varies faster  than the  sampling time.   It follows  that the
signal variance for double exposure time is $S_m(2\tau) = S_m(\tau) (1
+ \rho_m)/2$.

\begin{figure}
\centerline{
\includegraphics[width=8.5cm]{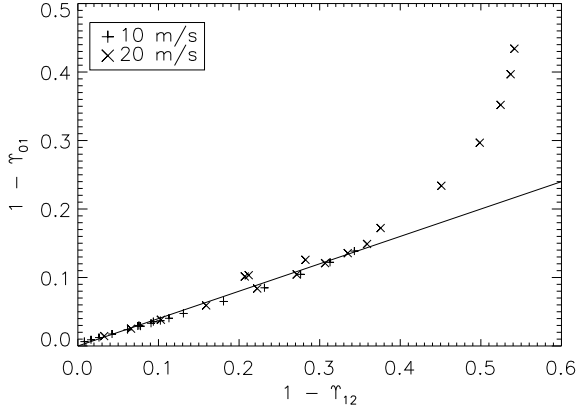} }
\caption{The signal attenuation  ratio for the double  and single exposure
  $\gamma_{12}$ is compared  to the signal attenuation  for the single
  and zero  exposure $\gamma_{01}$  for the spatial  blurs of  1\,cm (plus
  signs, wind  speed 10  m~s$^{-1}$ for 1-ms  exposure time)  and 2\,cm
  (crosses) and various  $m$.  The line is $1 -  \gamma_{01} = 0.4(1 -
  \gamma_{12})$.
  \label{fig:gamma} 
}
\end{figure}

Reduction of the signal variance for double exposure time $\gamma_{12}
= S(2\tau)/S(\tau) = (1 + \rho)/2$  (the index $m$ is omitted here) is
used to estimate the signal variance  with zero exposure, i.e.  the factor
$\gamma_{01}$.   The analytical  WFs  show  than  when  the  exposure-time
reduction is  not too severe (i.e.  when the blur $V  \tau$ is small),
the  two   factors  are  proportional   with  a  coefficient   of  0.4
(Fig.~\ref{fig:gamma}). The resulting correction  to the zero exposure
is
\begin{equation}
S_m(0) \approx S_m(\tau)/(0.8 + 0.2 \rho_m) .
\label{eq:S0}
\end{equation}
The correction  is very good  at $\rho  >0.2$ and still  acceptable at
$\rho  >0$, which  implies the  maximum exposure-correction  factor of
1/0.8=1.25.

The calculation shows  that with a 1\,ms exposure  time, the reduction
of the WFs can be substantial,  especially at large $m$.  For example,
at  $z=1$\,km  and  $V=10$\,m~s$^{-1}$,  the  correlation  coefficient
$\rho_m$ drops to  zero at $m=14$ and becomes negative  at larger $m$.
At  $z=16$\,km   and  $V=20$\,m~s$^{-1}$,  the  zero   correlation  is
encountered already  at $m=6$.  However, the  correlation coefficients
$\rho_m$ measured  so far  experimentally are above  0.6 for  all $m$.
Contrary to the  simulation, the fastest signals are  those with small
$m$. This  happens because  fast and high  atmospheric layers  are the
main contributors  to the  small-$m$ signals, while  at large  $m$ the
signal comes mostly from the slow turbulence closer to the ground.

The temporal variation  of the RINGSS signals is used  to estimate the
atmospheric time constant $\tau_0 = 0.31 r_0/\bar{V}$, where $\bar{V}$
is the turbulence-weighted effective wind speed. Although the standard
theory uses the  mean $V^{5/3}$ as a measure of  $\bar{V}$, the second
moment of the wind speed $V_2$ is more relevant for the performance of
AO systems and interferometers \citep{FADE}:
\begin{equation}
V_2^2 = J^{-1} \int C_n^2(z) V^2(z) {\rm d}z ,
\label{eq:V2}
\end{equation}
where $J  = \int C_n^2(z)  {\rm d}z$  is the turbulence  integral. The
corresponding  atmospheric AO  time constant  is $\tau_0  \approx 0.31
r_0/V_2$.   The effective  wind speed  $V_2$  does not  depend on  the
turbulence strength,  $r_0$, and $\lambda$,  and is a  more meaningful
parameter than $\tau_0$.

The difference between (\ref{eq:V2}) and (\ref{eq:stau}) is in the
factors $U$ under the integral. The idea of Kornilov is to combine
several U-functions with coefficients $C^U_k$ such that 
\begin{equation}
\sum_k C^U_k U_k(z) \approx 1,
\label{eq:CU}
\end{equation}
i.e. to remove  the dependence  on the  propagation distance  $z$.  Then
eq.~(\ref{eq:stau}) can be transformed to get
\begin{equation}
V_2^2 \approx \sum_m C^U_m \Delta_m,
\label{eq:V2est}
\end{equation}
where the right-hand side contains the measured quantities 
\begin{equation}
\Delta_m = 6 \frac{S_m(2\tau) - S_m(\tau)}{4\tau^2 - \tau^2} =
S_m(\tau)(1 - \rho_m)/\tau^2 .
\label{eq:Delta}
\end{equation}
The  coefficients  $C_m^U$  are   found  by  solving  eq.~\ref{eq:CU},
following the Kornilov's recipe.  In other words, we look for a linear
combination of  the U-functions  that is approximately  independent of
$z$.  The first tests indicated that signals  with $m$ of 2 and 4 have
very  small  coefficients  $C_m^U$,  while $a_6$  and  $a_7$  are  the
strongest contributors.   Therefore, the  set of U-functions  used for
the wind estimation in RINGSS is restricted to $m=[1,3,6,7,8,9]$.  The
corresponding coefficients  $C_m^U$ are $[9.8, 8.8,  20.2, 19.0, 17.5,
  14.6] \times  10^{-15}$ in  one representative case.   The resulting
linear combination of  the U-functions deviates from one by  less than 0.1
at $z>1$\,km  but falls  to zero  at the  ground.  Although  RINGSS is
sensitive  to the  turbulence near  the  ground, its  estimate of  the
effective wind speed refers mostly to the free atmosphere, as in MASS.
The  U-functions  and the coefficients  $C_m^U$ depend  strongly  on  the
spectral response and are re-computed together with the WFs.

\subsection{Saturation correction}
\label{sec:sat}

\begin{figure}
\centerline{\includegraphics[width=8.5cm]{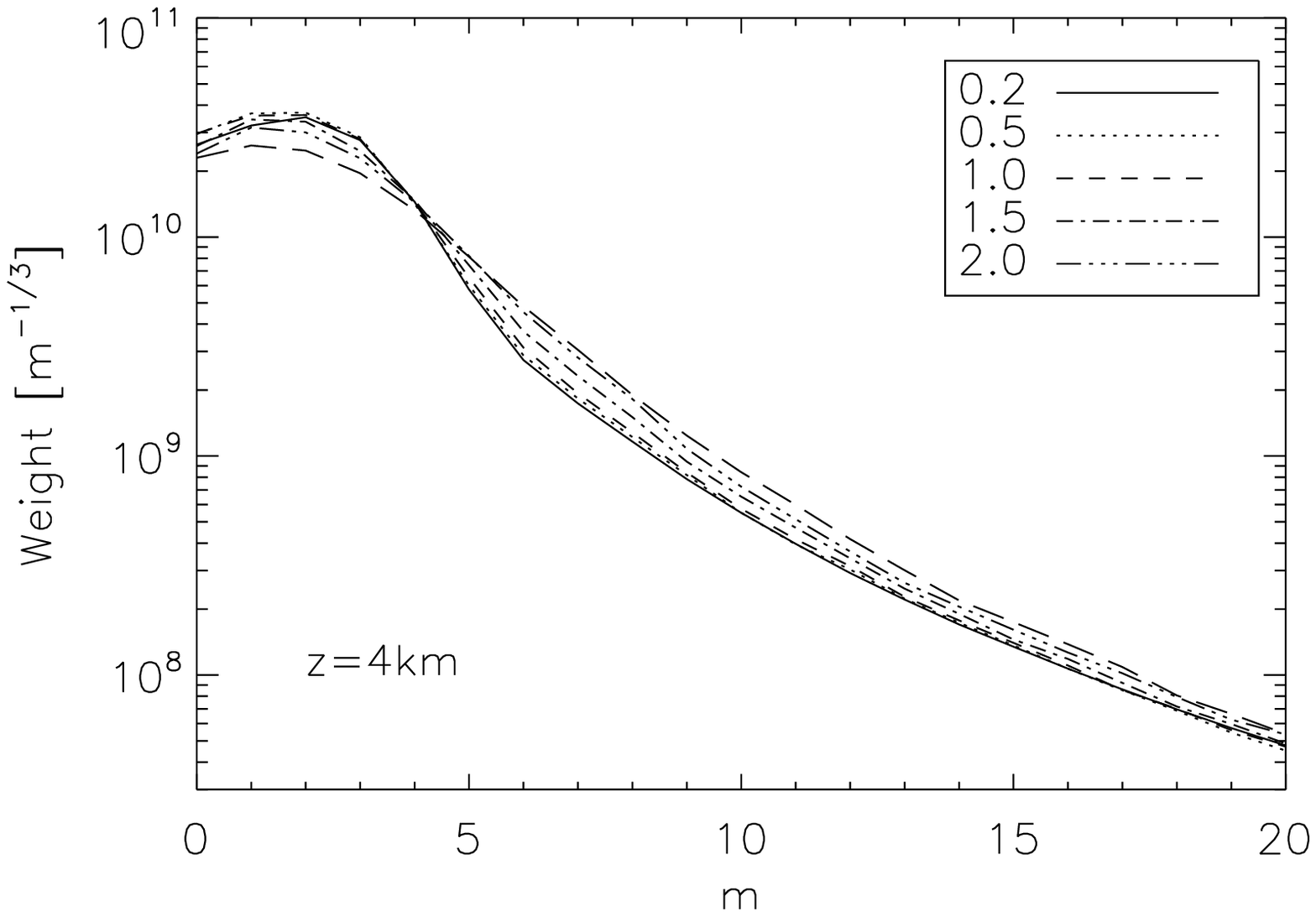} }
\centerline{\includegraphics[width=8.5cm]{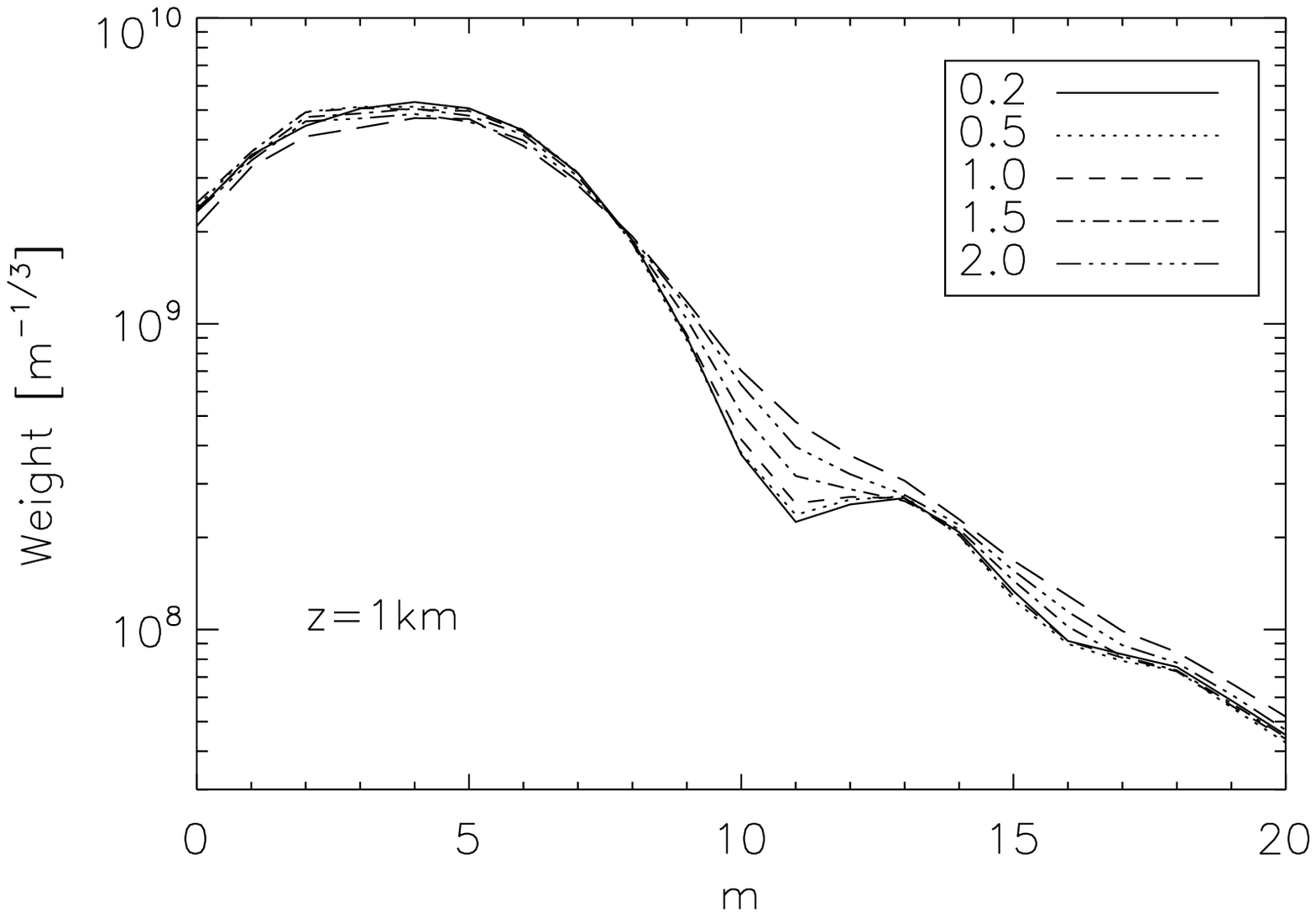} }
\caption{Comparison of  the analytic WFs  (solid line)  with the  WFs from
  numerical simulation  for seeing values  from $0.2''$ to  $2''$ (see
  the legend box)  and a single turbulent layer at  4\,km (top) and at
  1\,km (bottom).
  \label{fig:weights} }
\end{figure}

The  theory of  all  scintillation-based  turbulence sensors  (SCIDAR,
MASS, FASS, RINGSS) is based  on the small-signal (weak scintillation)
approximation that is not quite fulfilled in the real conditions.  The
spatial spectrum  of a  strong (semi-saturated)  scintillation differs
from the theoretical (weak-scintillation)  spectrum by containing more
high-frequency power  and less low-frequency  power. As a  result, the
APS  $S(m)$   increases  at   large  $m$,  imitating   a  low-altitude
turbulence,  and the  seeing  is  over-estimated (over-shoots).   This
effect was  extensively studied in  the case  of MASS, and  its partial
correction   based  on   numerical   simulations   was  developed   by
\citet{Tok2007}.  The  idea is  to transform the  measured APS  to the
values that would be obtained without saturation and then to apply the
standard  linear  profile  restoration algorithm.   This  strategy  is
studied here for RINGSS, and it  is also applicable to the pupil-plane
sensors like FASS.

Figure~\ref{fig:weights} compares the theoretical (weak-scintillation)
weighting  functions  (WFs)  with   the  results  of  simulations  for
monochromatic light  to illustrate  the impact of the increasingly strong
scintillation. For a turbulent layer at 4\,km, the classical effect is
observed, namely  decrease of the power at low frequencies (small $m$)
and its  increase at  intermediate frequencies, $m>5$.  The cross-over
occurs at  $m=4$.  Note that at  $m \sim 20$ the  impact of saturation
becomes smaller.  For a layer  at 1\,km, the overall  scintillation is
smaller, and the effect of saturation is moderate below the cross-over
at  $m=8$.  However,  the  WF   minimum  at  $m=11$  is  progressively
filled.  If the  effect  of  saturation is  expressed  by  the ratio  of
the simulated  and theoretical  WFs, there  is a  strong spike  around
$m=11$, reaching a factor of  two.  Summarizing, there are two distinct
effects  of  saturation:  (i)   progressive  transfer  of  power  to
intermediate frequencies at large $z$  and (ii) partial filling of the
WF minima at small $z$. The second effect is specific to RINGSS and is
not present  when the scintillation  is measured  at the pupil;  it is
presumably  caused  by  the  interplay  between  the phase  and  amplitude
distortions. This second effect, however, can be neglected in practice
because at those frequencies the WFs are an order of magnitude smaller
compared to their maximum.

The strength of the scintillation is  characterized by the total intensity
variance (the Rytov number) $s_0^2$ \citep{Roddier81}
\begin{equation}
s_0^2 = 19.12 \lambda^{-7/6} \sum_j J_j z_j^{5/6} .
\label{eq:rytov}
\end{equation}
The weak-scintillation regime is valid  at $s_0^2 \ll 1$, while $s_0^2
\sim 1$  is the regime  of strong (semi-saturated)  scintillation. This
situation  is sometimes  encountered in  practice.  The  maximum Rytov
variance for the plots in  Fig.~\ref{fig:weights} is 0.80 and 0.25 for
$z=4$\,km and  $z=1$\,km, respectively.  The parameter  $s_0^2$ is not
measured directly,  but the sum  of APS, $S_{\rm tot}  = \sum_m S_m$,  is a
valid  substitute  because   the  proportionality  $S_{\rm  tot}/s_0^2
\approx 0.4$ holds according to the simulations.

A large set of monochromatic simulations was carried out to develop an
approximate correction of  semi-saturated scintillation, following the
MASS prescription \citep{Tok2007}. Each simulation involves two layers
with distances ranging from  0.5\,km to 16\,km,  selected randomly
  from   a  fixed   logarithmic   distance  grid   with   a  step   of
  $\sqrt{2}$.  Turbulence integrals  $J$  in these  layers are  chosen
  randomly  to  produce   the  Rytov variance  from  0.05  to 1.  The
measured APS $S_m$ is compared to the theoretical APS $S_m^{\rm theo}$
calculated  from  the  WFs,  i.e.  corresponding  to  the  unsaturated
regime. Their ratio is approximated by the formula
\begin{equation}
S_{m}/S_m^{\rm theo} \approx 1 + \sum_{k=1}^{5} Z_{m,k} S_k, 
\label{eq:cor1}
\end{equation}
where the sum includes a restricted number of terms from $k=1$ (the
$k=0$ term is not used) to 5. The translation of the measured APS to the
quasi-linear one is the inverse of the right-hand part; it
approaches one when the scintillation tends to zero.

The set  of linear  equations (\ref{eq:cor1}) is  solved by  the least
squares  method  to  find   the  coefficients  $Z_{m,k}$,  called  
Z-matrix.   Two  subtleties  are  relevant.   First,  only the simulation
results with the  Rytov variance from 0.05  to 0.7 are used  and the cases
with  the  lowest layer  at  or  below  1\,km  are excluded  from  the
`training  set' used  to find  $Z$.   Second, the  inversion of  the
system matrix of  the least-squares problem is done  by the singular value
decomposition with rejection of singular values below $10^{-3}$ of the
largest singular value, to  avoid the noise amplification.  Typically,
3 singular values are rejected.

\begin{figure}
\centerline{
\includegraphics[width=8.5cm]{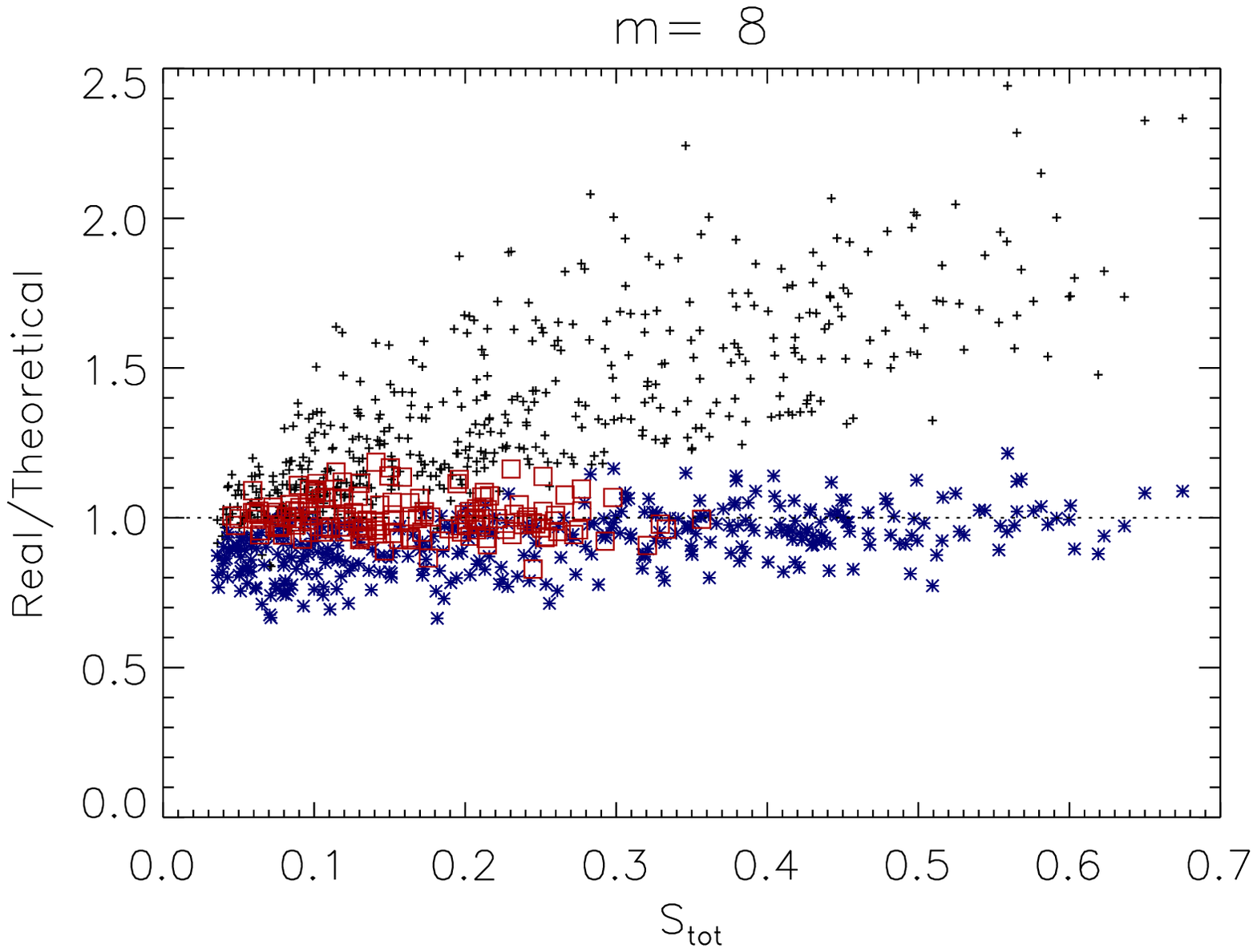} }
\centerline{\includegraphics[width=8.5cm]{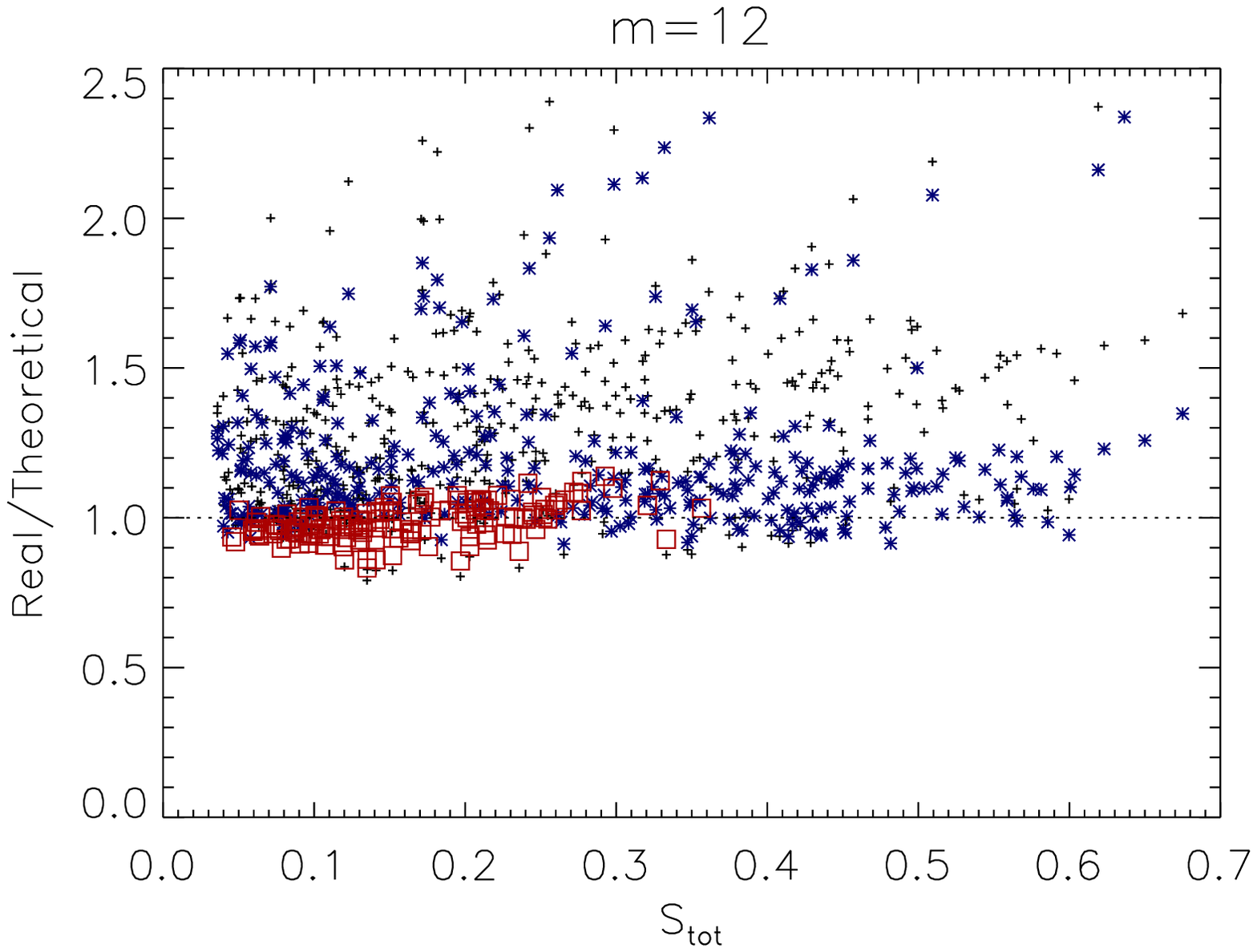} }
\caption{The  ratio  of  the  uncorrected  to  the  theoretical  power
  $S_m/S_{m}^{\rm theo}$  (crosses), the ratio of  the corrected power
  for  the training  set (red  squares), and  the same  ratio for  the
  remaining cases  (blue asterisks) are plotted  against $S_{\rm tot}$
  for $m=8$ (top) and for $m=12$ (bottom).
  \label{fig:testzmat} 
}
\end{figure}

The matrix correction was determined for all $m$ from 1 to 20 that are
used in the profile  restoration.  Figure~\ref{fig:testzmat} gives two
representative plots.  The  quality of the correction  is estimated by
the  rms of  the  ratio $S_{m}^{\rm  corr}/S_{m}^{\rm theo}$  computed
separately for  the training set and  for the full set.   Overall, the
correction works  quite well, and  the rms residuals for  the training
set are between  0.05 and 0.09, depending on $m$.   The largest impact
of the saturation  and, correspondingly, the largest  correction, is found
for $m=9$, and the smallest one for $m=1$ and $m=2$.

Looking  at  Fig.~\ref{fig:testzmat}, one  notes  that  for $m=8$  the
correction works very well not only for the training set, but also for
the full set;  the rms for the  full set is within  0.1.  However, for
$m=12$ some blue  asterisks are well above one, while  the rms for the
full set, 0.23,  is almost as large as 0.31  without correction. These
deviant points correspond to the  cases with layers below 1\,km, where
the  minima  of  the   WFs  are  filled  (see  Fig.~\ref{fig:weights},
bottom). This phenomenon is not corrected by the current algorithm and
for this reason  the low-$z$ cases are removed from  the training set.
The mixture of two different phenomena related to strong scintillation
complicated the development of the saturation correction algorithm for
RINGSS.

\begin{figure}
\centerline{\includegraphics[width=8.5cm]{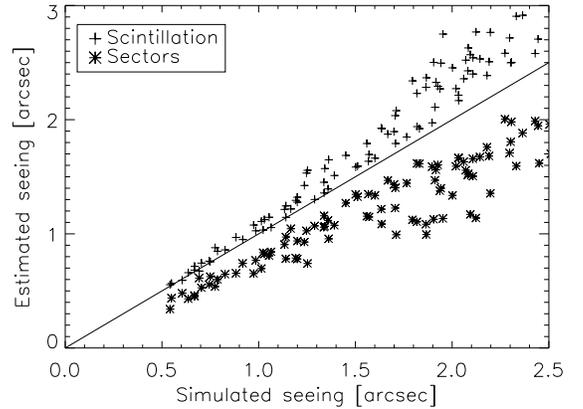} }
\caption{Processing of  the simulated image cubes:  the seeing deduced
  from the  turbulence profiles  restored from  the APS  (crosses) and
  estimated from the differential sector motion (asterisks) is plotted
  against  the   true  simulated   seeing,  with  the   straight  line
  corresponding to equality. The saturation is not corrected.
  \label{fig:seecorr} }
\end{figure}

Another, smaller set of monochromatic  simulations was made to emulate
1-second  image cubes  and to test the  saturation correction.   Without
correction, the  seeing deduced from the  restored turbulence profiles
is systematically  over-estimated, while the seeing  computed from the
differential       sector        motion       is       under-estimated
(Fig.~\ref{fig:seecorr}).   The  `overshoots'   are  caused  by  the
partially  saturated scintillation.  The  energy spilled  to a  higher
spatial frequency is  interpreted as coming from a  lower but stronger
layer, and the seeing is  over-estimated.  The restored OTP is shifted
to a  smaller $z$, compared to  the input OTP. These  effects are also
observed in MASS.

The   seeing  deduced   from   the  differential   sector  motion   is
under-estimated (under-shoots) for two  reasons. First, the decline of
the response  coefficient $C_r(z)$  at large propagation  distances is
not taken into account because the distance to turbulence is not known
a priori without  measuring the OTP.  This effect is  corrected for by
applying  the profile-weighted  coefficient in  eq.~\ref{eq:seeing} to
compute the seeing.   Secondly, even after this  correction, the seeing
estimate declines with increasing scintillation approximately as $ 1 -
0.78  S_{\rm tot}$.  The division  by this  empirical  factor brings  the
seeing estimated  from the  differential sector motion  into agreement
with  the  input  seeing.   Saturation  correction  also  removes  the
overshoots,  and,  after the correction,  both pluses  and  asterisks  in
Fig.~\ref{fig:seecorr} align  nicely along the diagonal  (this trivial
plot is not reproduced here).

The Z-matrix correction is a semi-empirical solution to the turbulence
profile restoration  for a  moderately saturated  scintillation.  This
regime is frequently encountered in practice, and the correction seems
to  be necessary;  otherwise,  the seeing  and free-atmosphere  seeing
become  over-estimated.   However,   these  overshoots  remain  modest
(mostly within 10 per cent) and can be considered as tolerable, especially at
good sites.  After all, turbulence parameters are always measured with
a varying  degree of  approximation.  The latest  version of  the MASS
data  reduction  software  \citep{KK2011}  does not  correct  for  the
saturation and is prone to overshoots.

The  phenomenon  of  saturated   scintillation  is  generic,  but  the
correction matrix  depends on  the instrument  parameters.  A  tool to
estimate  the Z-matrix  from simulations  for an  arbitrary instrument
will be developed by adaptation  of the existing code.  Qualitatively, the
effect of  saturation (decrease at  small $m$ and overshoots  at large
$m$) remains the same for any instrument; it was first noted in MASS.

\subsection{Turbulence profile restoration}
\label{sec:prof}

The turbulence profile is  modeled by 9 layers at fixed  distances, with 8
layers on  a log-spaced grid  at 0.125,  0.25, \ldots, 16\,km  and the
first layer at the ground. This crude model is similar to the one used
in MASS, except the additional low layers.  Turbulence located between
the fixed layers is attributed to the adjacent layers, while its total
strength is estimated with a  relative error not exceeding 10 per cent. Hence,
the crudeness of  the distance grid does  not contribute substantially
to the errors. The $z$-grid can be optimized in the future.

The OTP, i.e. the 9 values  of $J_j$, is the least-squares solution of
eq.~\ref{eq:wf}.   To avoid  negative  $J_J$,  the non-negative  least
squares method is used, as done by \citet{KK2011} for MASS. The system
of  equations is  over-determined when  more than  9 $S_m$  values are
used. The  estimated noise bias  is subtracted from the  measured APS.
The $m=0$ term (the full variance) is  not used because it is not well
measured from the short data cubes, being dominated by the slow signal
variation,  and potentially is affected  by  variable transparency.   The
high-order terms  with $m>20$ are  also not used because  they contain
little useful  signal. The exposure-time  correction (eq.~\ref{eq:S0})
is applied, and  the range of $m$ is further  restricted to terms with
non-negative  correlation coefficients  $\rho_m$, but  always includes
terms up to  $m=9$. This restriction was never necessary  for the real
data,  where all  20  $S_m$ values  from $m=1$  to  $m=20$ are  fitted
because they  have positive $\rho_m$.   Apart from the  exposure time,
the APS  values are  corrected for the  saturation using  the Z-matrix
(eq.~\ref{eq:cor1}).

\begin{figure}
\centerline{
\includegraphics[width=8.5cm]{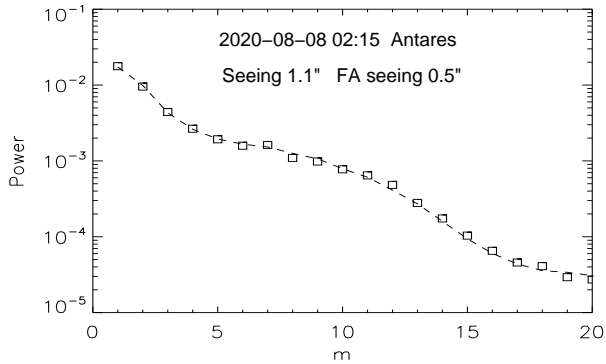} }
\caption{Example of  the OTP fitting to  the APS computed from  a real
  2-s data  cube.  Squares  -- measured APS,  dashed line  -- fitted
  model.
  \label{fig:profrest} 
}
\end{figure}

Figure~\ref{fig:profrest}  gives an  example  of  the OTP  restoration
using  real  data.  A  typical relative  rms  difference  between  the
measured  APS  and  its   model  is  $\sim$0.1.   Turbulence  profiles
determined from the successive 2-s records show a similar structure.

The  profile restoration  is  accompanied by  the  calculation of  the
effective  wind  speed  $V_2$  using the  pre-computed  set  of  $C^U$
coefficients   and  applying   eq.~\ref{eq:V2est}.   The   wind  speed
determined from the simulated data  matches the input speed, while the
real data show a good consistency between successive $V_2$ estimates.

The total seeing  is initially estimated from  the differential sector
motion by eq.~\ref{eq:seeing} using the coefficient $C_r$ computed for
$z=0$; the  4 estimates  of the differential  variance $\sigma^2_{2r}$
from  the  8   sectors  are  averaged  and  the   estimated  noise  is
subtracted. After the OTP is determined, these estimates are corrected
for  the propagation  and for the  saturation.  The  two estimates  of the
total seeing obtained  independently, one from the APS  (i.e. from the
sum of $J_j$) and another  from the differential sector motion, should
agree  mutually,  providing  an  overall control  of  the  measurement
procedure and  biases.  Figure~\ref{fig:seecorr} illustrates  the lack
of such agreement when the scintillation  is strong and the biases are
left uncorrected.

\section{Prototyping}
\label{sec:proto}

\begin{figure}
\centerline{\includegraphics[width=8.5cm]{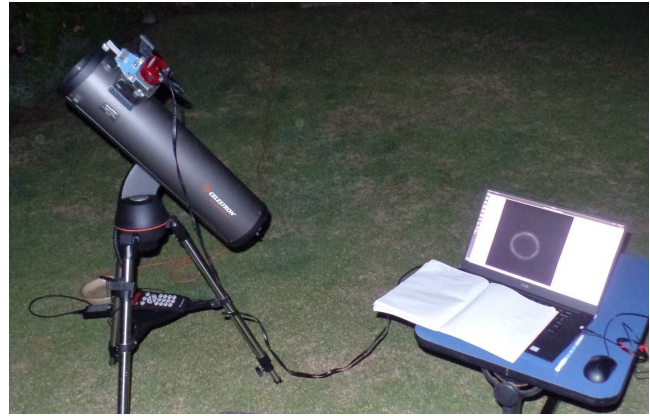} }
\caption{Photo of the RINGSS prototype.
\label{fig:proto} }
\end{figure}

This study is devoted mostly to  the algorithms and software needed for
the turbulence  measurement from  ring  images.   However, testing  these
algorithms on real data is essential  to prove that they actually work
and to find  potential caveats.

A  crude   prototype  RINGSS  instrument   (Fig.~\ref{fig:proto})  was
assembled  based on  the Celestron  Nexstar SLT-130  reflecting Newton
telescope with  $D=0.13$\,m and central obscuration  $\epsilon =0.45$,
defined  by  a mask  behind  the  secondary  mirror (other  values  of
$\epsilon$ were also  tested).  The spiders holding  the secondary mirror are
so  thin that  their shadows  are not  seen in  the ring  images.  The
effective focal distance was shortened  twice, from the native 0.65\,m
to  0.35\,m,  by placing  an  achromatic  doublet  lens with  a  focal
distance of $f=50$\,mm at $f/2$ distance  in front of the camera.  The
lens also  provides a spherical  aberration needed to get  sharp rings
for a defocus of 1.06\,mm that corresponds to the conjugation distance
of $H=400$\,m.  A yellow filter cuts  off the blue light short-ward of
$\sim$450\,nm.

The detector is a monochrome CMOS  camera ZWO ASI290MM used mostly for
amateur
astrophotography.\footnote{https://astronomy-imaging-camera.com/product/asi290mm}
The pixel  size is  2.9 $\mu$m,  format 1936$\times$1096  pixels (size
5.6$\times$3.2  mm),   readout  noise  $\sim$1  el,   maximum  quantum
efficiency (QE)  0.80. The  detector is  a back-illuminated  CMOS chip
IMX290 from Sony, uncooled. The high QE and the low noise specified by
the vendor were confirmed by  our independent characterization of this
camera.  The  conversion factor from  ADU to electrons (in  the 12-bit
mode) is $3.6  \times 10^{-G/200}$, where $G$ is the  camera gain setting.
With $G=300$, the  conversion factor is 0.11 el/ADU and the  readout noise is
1.0 el in  most pixels (a 3$\times$  larger noise is found  in a 0.2 per cent
fraction  of  pixels,  which  is typical  for  other  scientific  CMOS
cameras).  The spatial uniformity is very  good, with a pixel to pixel
sensitivity variation of 0.3 per cent rms.

\begin{figure}
\centerline{ \includegraphics[width=8.5cm]{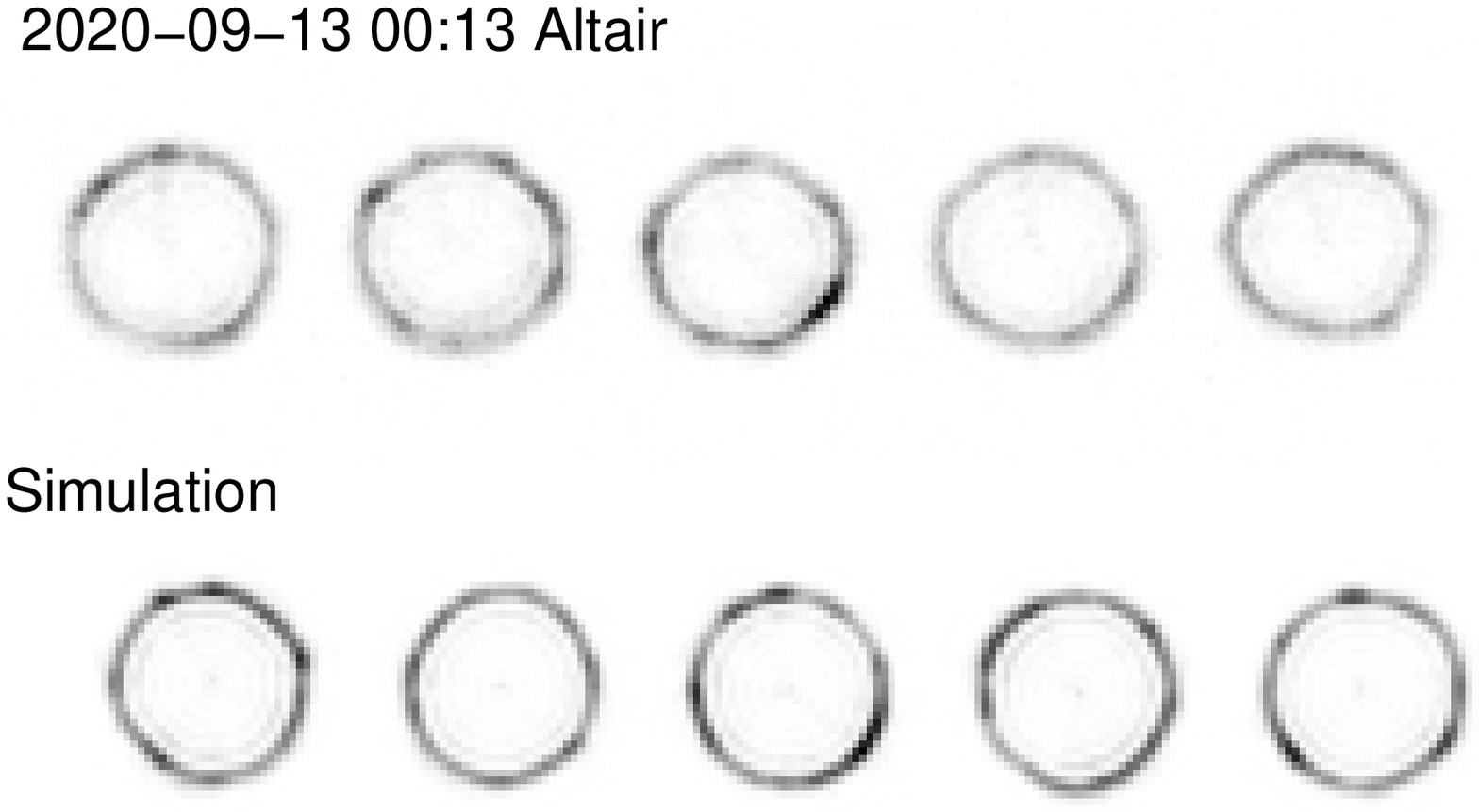} }
\caption{Top:  five ring-like  64$\times$64 images  recorded with  the
  prototype  instrument (ring  radius  13.1  pixels, estimated  seeing
  $1.6''$).   Bottom: five  images  from a  simulated  data cube with
  similar parameters.
  \label{fig:cubes} 
}
\end{figure}

The pixel scale of the prototype is $1.7''$, so the full field of view
is about 0.5$^\circ$; the typical ring  radius is 12 pixels or $20''$.
Image cubes of 64$\times$64$\times$2000 pixels with 1\,ms exposure per
frame  were acquired  using the  {\tt  ASICap} software  from ZWO.   A
stand-alone    acquisition    software    for   Linux    is    under
development. Figure~\ref{fig:cubes}  compares a  series of  the actual
ring      images     with      simulated      rings     (see      also
Fig.~\ref{fig:imagecompare}).

\begin{figure}
\centerline{\includegraphics[width=8.5cm]{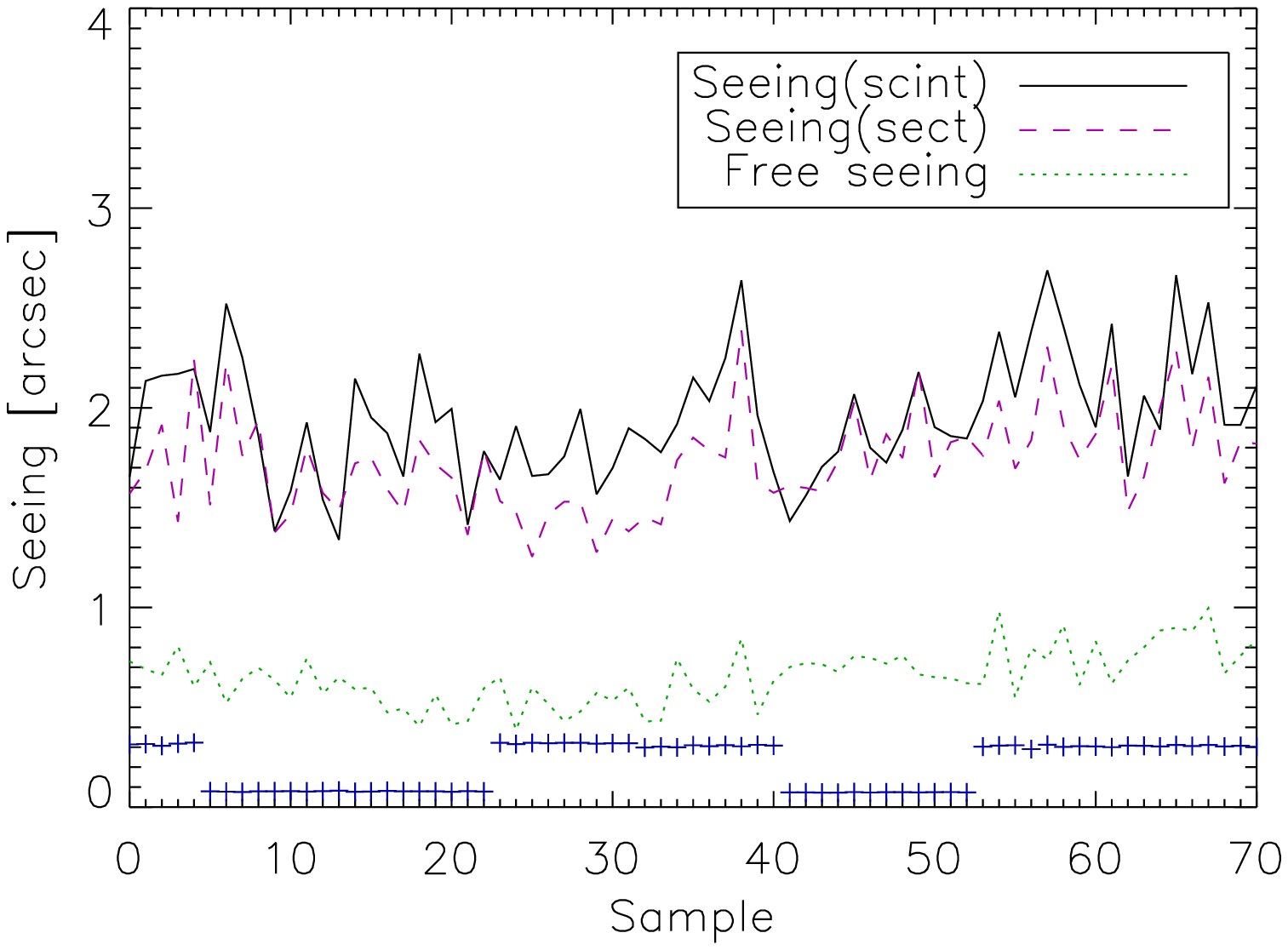} }
\centerline{\includegraphics[width=8.5cm]{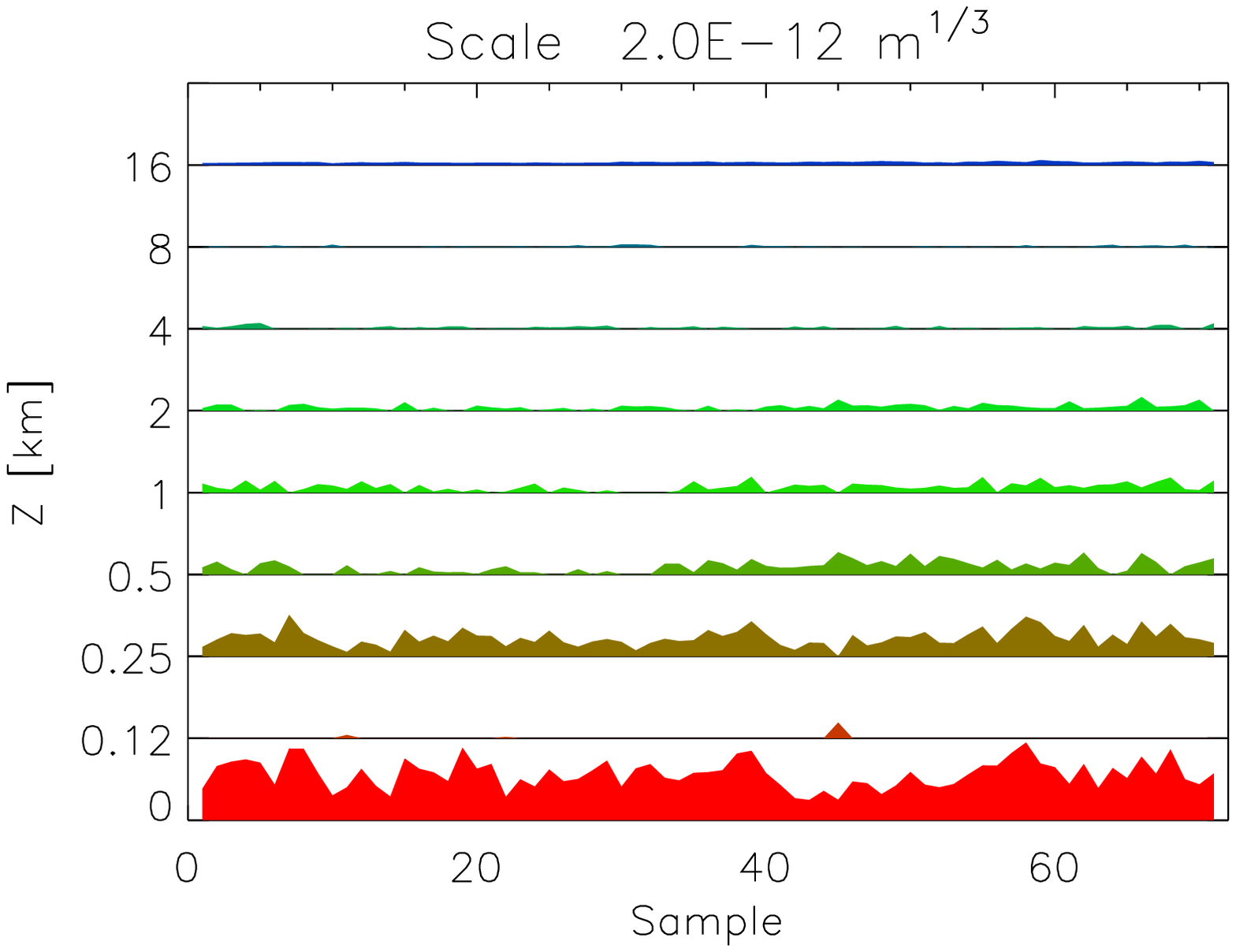} }
\caption{Data from  the prototype instrument taken  on 2020-09-12. Top
  panel: the total seeing estimated from the OTP (solid line) and from the
  differential sector  motion (dashed  magenta line),  as well  as the
  free-atmosphere seeing above 0.5\,km (dotted green line), vs. sample
  number. Blue crosses indicate the flux in  arbitrary units to show the change
  of stars.  Bottom panel: OTPs for  the same night; the width of each
  band corresponds to $J = 2 \times 10^{-12}$\,m$^{1/3}$.
\label{fig:sep12} }
\end{figure}

Test data  were acquired with  this prototype at  the sea level  in La
Serena, Chile (access  to the Cerro Tololo observatory  was closed due
to  the COVID-19  pandemic). The  seeing is  poor by  the astronomical
standards, but some  data with a moderate level  of scintillation were
nevertheless    obtained.     One    example   is    illustrated    in
Fig.~\ref{fig:sep12}.  The  data were  taken during two  short periods
separated by  2 hours (the second  set begins at the  sample 32).  The
conditions  were   stable,  the   estimated  wind  speed   was  around
6\,m~s$^{-1}$, and  the scintillation was small,  $S_{\rm tot} \approx
0.1$.  During each period, the  instrument was pointed to $\alpha$~Aql
($V$=0.76 mag, spectral type A7V), then to $\gamma$~Aql ($V$=2.70 mag,
K3II)  and back.   Crosses in  Fig.~\ref{fig:sep12} indicate  the star
change.  The average flux from these stars in 1-ms exposures was $3.1 \times
10^4$ and $7.7 \times 10^3$  el, respectively.  Despite the flux variation
by a factor of  4, the change of star does  not seem to affect
the measured seeing. The WFs were calculated for the effective temperature
of 7500\,K  corresponding to  $\alpha$~Aql.  The mismatching  color of
$\gamma$~Aql does not  seem to affect the  results strongly.  Note that
each measurement is obtained from the image cube of only 2 s duration.

\section{Instrument parameters}
\label{sec:instr}

In this Section, some considerations on the RINGSS instrument  are
given. The parameters of the prototype are adequate for turbulence
monitoring, but are they chosen optimally? 

\begin{table}
  \center
  \caption{Main instrument parameters}
  \label{tab:par}
  \begin{tabular}{ll}
    \hline
    Parameter  & Formula \\
    \hline
    Ring radius (rad) & $r_{\rm ring, rad} = D(1 + \epsilon)/(4H)$ \\
    Ring radius (pixels) & $r_{\rm ring} = D(1 + \epsilon)F/(4Hp)$ \\
    Ring thickness (rad) & $\delta r = 2 \lambda/[D (1 - \epsilon)]$ \\
    Pixel scale (rad)    & $p/F$ \\
    Optimum focal ratio & $F/D \approx p(1 - \epsilon)/\lambda$ \\
    Intrafocal distance (m) & $\Delta = F^2/H$ \\
   \hline      
  \end{tabular}
\end{table}

The main parameters (telescope diameter $D$, focal length $F$, central
obscuration $\epsilon$, conjugation distance  $H$, and pixel size $p$)
are mutually  related.  The formulae are  given in the text  above and
repeated in Table~\ref{tab:par} for  convenience. If the optimum $F/D$
is selected to  sample the ring width  by 2 pixels, this  leads to the
formula for  the ring  radius in  pixels that does  not depend  on the
physical pixel size:
\begin{equation}
r_{\rm ring} \approx D^2 (1 - \epsilon^2) /(4 \lambda H)  .
\label{eq:rpix}
\end{equation}
For a fixed 2-pixel ring width, the total number of illuminated pixels
is proportional to  $r_{\rm ring}$, hence to $D^2$. The  total flux is
also proportional to $D^2$, so the number of photons per pixel remains
constant if  $\epsilon$ and $H$  do not  change. However, in  order to
limit the  size of the  image cubes, it  is advisable to  increase $H$
when a larger $D$ is chosen. 

A larger aperture collects more photons, produces sharper ring images,
and thus substantially reduces the  noise of differential image motion
measurement.  Implementation  of the  RINGSS concept with  $D =  0.2 -
0.3$\,m is certainly  feasible, although it requires a  larger size of
the image  cubes (see above).   The Kolmogorov turbulence  spectrum is
dominated by the  large-scale features; as a result,  the amplitude of
the wavefront  distortions increases  as $D^{3/5}$  and the  signal to
noise ratio in an instrument like RINGSS, DIMM, or FADE also increases
with $D$.   However, the use  of the full  aperture in RINGSS  and the
modern light detectors with a low noise and a high QE allow us to measure
the  seeing   with  smaller   apertures  (compared  to   DIMM)  with
 a  reasonably low noise.  Calculation  indicates that with
$D=0.1$\,m,  the  noise of  the  differential  sector motion  using  a
$V=2.5$ mag star is equivalent to a $0.16''$ seeing, so the noise bias
is still  moderate and  easily accounted  for.  At  $D <  0.1$\,m, the
noise  increases quickly  and the  seeing measurement  by differential
sector motion becomes questionable, unless  very bright stars are used
and the seeing itself is large.   The aperture of $D \sim 0.1$\,m also
matches the  Fresnel radius of  high-altitude turbulence and  is large
enough for OTP measurement by scintillation in MASS, FASS, or RINGSS.

Calculations  show that  the ring  thickness is  minimum at  $\epsilon
\approx 0.45$, and  such central obscuration is  optimal for measuring
the  differential sector  motion. A  relatively wide  aperture annulus
averages the small-scale scintillation, reducing  the WFs at large $m$
in comparison  with a narrower (larger  $\epsilon$) annulus.  However,
the increase in the photon flux almost compensates for the loss of the
scintillation  signal, so  the central  obscuration of  $\epsilon \sim
0.5$ is good for measuring both the differential sector motion and the
scintillation.   In  FASS, the  scintillation  is  measured in  narrow
annular zones chosen  inside the aperture (or in  the defocused image)
to  avoid spatial  averaging,  while in  RINGSS  the scintillation  is
averaged in the radial direction.

\begin{figure}
\centerline{\includegraphics[width=8.5cm]{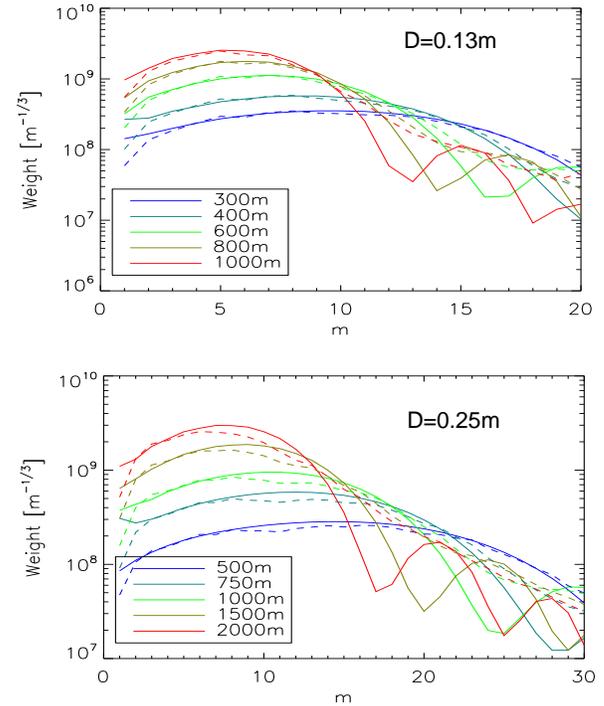} }
\caption{Weighting  functions   for  turbulence  at  the   ground  and
  different conjugation distances  $H$ (see the legend  box). The full
  lines are analytical  calculations, the dashed lines  are derived by
  simulation  assuming  a  $2''$  seeing and the monochromatic  light  of
  $\lambda  =  600$\,nm.  The  top  plot  refers  to  an  aperture  of
  $D=0.13$\,m, the lower plot to $D=0.25$\,m, both for $\epsilon = 0.5$.
\label{fig:testh}}
\end{figure}

The choice of the conjugation distance $H$ (or, equivalently, the ring
radius) is  driven by the compromise.  A small $H$ (large  ring) gives
access to the small-scale scintillation (large $m$), but its amplitude
is  also   quite  small.  By   increasing  $H$,  we  get   a  stronger
scintillation from the ground layer and a larger number of photons per
pixel in a smaller ring. By setting the minimum ring radius of $r_{\rm
  ring}=10$ in (\ref{eq:rpix}) and assuming $\epsilon=0.5$, we
get $D = 7.3 \sqrt{\lambda H}$, or 7 Fresnel zones across the pupil for
the propagation distance $H$. So, to keep a fixed ring radius of 10
pixels, we need to increase $H$ in proportion to $D^2$: 0.5\,km for
$D=0.13$\,m and 2\,km for $D=0.25$\,m.

Strong    turbulence    is     typically    encountered    near    the
ground. Fluctuations  of the  ring images  caused by  the ground-layer
turbulence   can   also   be  strong,   violating   the   small-signal
approximation. This problem is investigated in Fig.~\ref{fig:testh} by
numerical  simulation for  the case  of very  poor $2''$  ground-layer
seeing.  At  small  $H$  (blue   lines),  the  agreement  between  the
analytical and simulated  WFs is quite good, but it  becomes worse for
the larger $H$, especially at large $m$.  The minima of the WFs become
progressively   filled,   which    is   typical   for   semi-saturated
scintillation (compare with Fig.~\ref{fig:weights}). At the same time,
the WFs increase with $H$.

Resuming, the choice of $H \sim  0.5$\,km for our prototype appears to
be  optimal. A  smaller $H$  results in the decreased sensitivity  to the
ground-layer turbulence and in a lower number of photons per pixel, while
for a  larger $H$  the ring radius  would be too  small. A  larger $H$
should be selected  for a larger aperture. A good  compromise might be
to select $H  \propto D$, in which case $r_{\rm  ring} \propto D$
also. The exact value of $H$ is not critical as long as it is not very
different from the optimum. However,  $H$ must be accurately known for
correct  calculation of  the  WFs. In  practice,  the telescope  focus
should be controlled to keep the ring radius at the desired value.

Keeping  a constant  defocus (hence  a constant  ring radius)  is also
important  for getting  sharp  (diffraction-limited)  ring images.  As
noted  above,  a  conic  wavefront  is  obtained  when  the  spherical
aberration and  defocus have opposite  signs and their  rms amplitudes
are in a  1:10 proportion. The spherical aberration is  defined by the
optics, so this optimum ratio corresponds to the specific value of the
defocus. Once  the $H$  is chosen,  the optical  design of  the system
should provide the required amount of the spherical aberration. In the
prototype, this  is achieved by  choosing the focal-reducing  lens and
the de-magnification factor. Suitable commercial achromatic lenses can
be found  to provide  the desired amount  of spherical  aberration and
focal-distance reduction  for almost  any feeding telescope,  but some
residual chromatic aberration might be  present as well. A custom lens
design  would  solve  this  problem. Instead  of  using  a  commercial
telescope,  RINGSS can  be  fed  by a  single  concave  mirror with  
suitable focal distance and  conic constant, delivering achromatic ring
images without any lenses.

The optics  of a real  instrument is  never perfect.  The  most likely
aberrations are astigmatism  and coma. I studied  biases introduced by
the coma and found  that a coma up to 2 radian  rms is acceptable. The
absence of  coma is readily controlled  by the uniformity of  the ring
images (see Fig.~\ref{fig:imagecompare}).  Careful optical alignment
of  the feeding  telescope and  the  stability of  this alignment  are
required for RINGSS.

\section{Discussion}
\label{sec:disc}

The  new  concept of  a  site  monitor  emerging emerging from this  study  is
attractive  for several  reasons.   First, its  hardware  is based  on
readily available commercial components and is relatively inexpensive,
within  the  reach  of  an amateur  astro-photographer.   Second,  the
telescope aperture  needed to monitor  the seeing is reduced  from the
typical 20-30\,cm to 10-15\,cm. A  smaller telescope can use a smaller
mount and can be housed in a smaller enclosure; the cost reduction and
the increased portability are evident.  The relatively large pixels of
RINGSS and the short exposure time  reduce its sensitivity to the wind
shake, compared to a standard DIMM.

While  the  hardware is  simplified,  more  burden  is placed  on  the
software.   Its  main  component  is  the  image-processing  tool  for
measuring the angular  intensity variation in the ring  images and the
fluctuations of  their radii. The  second key  element is the  tool to
compute the  WFs that depend on  the instrument parameters and  on the
spectral  response.  Once  the WFs  are known,  the estimation  of the
turbulence  profile  is  trivial,  but,  as  in  any  seeing  monitor,
attention should be paid to the correction of biases.

Are the results of RINGSS reliable?  This is the fundamental issue for
any turbulence monitor.  Scintillation sensors are `self-calibrated'
because intensity  fluctuations that  they record  are related  to the
turbulence parameters  by a well-established  theory, at least  in the
weak-scintillation regime.  In the case  of MASS, departures from this
regime must  be corrected  for, and accurate  account of  the spectral
response  is needed;  both issues  are equally  relevant for  FASS and
RINGSS. Moreover, RINGSS measures the signal close to the image plane,
not  at  the  pupil.   This  makes little  difference  for  a  distant
turbulence,  but a  large difference  for the  near-ground turbulence.
However, RINGSS also estimates the seeing by the alternative method of
the  differential sector  motion that  is a  variant of  a DIMM.   The
agreement between these two independent estimates of the same quantity
gives  some assurance  that  both are  correct.   A similar  agreement
between MASS and  DIMM is observed when  turbulence is predominantly
high; for  RINGSS such agreement should  always hold and is  a sign of
the correct  data processing.  The  results of RINGSS are  anchored to
the  turbulence  theory  to  the  same extent  as  they  are  for  the
alternative turbulence monitors, and in  this sense they are reliable.
Experimental comparison between  turbulence monitors, preferably based
on  different principles,  is  a  useful way  to  check their  biases.
However,  the idea  of `calibrating'  one monitor  against another  is
misleading  because none  gives  totally unbiased  results, while  the
biases depend on many factors, rendering such calibration meaningless.
A recent comparison campaign of FASS is reported by \citet{FASS2}.

The  number of shortcuts  and approximations  needed to  interpret the
signal of  RINGSS is  impressive. However, similar  approximations are
involved  in any  turbulence monitor  (although not  always recognized
explicitly), and  the atmospheric  theory itself is  only approximate.
This point is further discussed by \citet{Tok2007} in relation to DIMM
and MASS.   To give  an example, the  `golden standard'  of turbulence
profiling,  SCIDAR,  uses analytical  monochromatic  WFs that  neglect
pixel and exposure-time averaging   \citep[however, see the recent
    paper  by][]{Butterley2020},   while  the  deviations   from  the
weak-scintillation regime are also ignored \citep{Osborn2018}.

The ability to measure the total seeing, a crude turbulence profile, and the
atmospheric  time  constant is  all  that  is  needed for  a  portable
site-testing  turbulence  monitor.   RINGSS  is  developed  with  this
application in mind. It can also serve as a regular turbulence monitor
at  the   existing  observatories,   replacing  the   aging  MASS-DIMM
instruments.

The RINGSS concept  is flexible. It can be applied  to both larger and
smaller  apertures.  Existing  DIMMs  can  be  converted  into  RINGSS
turbulence  profilers by  replacing their  cameras and  software.  The
concept will also work for defocused images without radial sharpening;
however, the  number of  photons per  pixel would  be reduced  and the
alternative seeing measurement by the differential sector motion would
be lost.   RINGSS with a detector  conjugated to the pupil  becomes an
incarnation of FASS with only a minor difference (the annular aperture
is integrated radially in RINGSS but split into narrow rings in FASS).

Historically, the differential  image motion was, on  one hand, easily
related  to  the   seeing  theoretically  and,  on   the  other  hand,
technically feasible to measure in the 1980-s and later (at that time,
detectors  and  computers for  fast  recording  of scintillation  were
either not  available or  complex).  The development  of MASS  was enabled
technically by  the progress  in electronics  and required  a matching
effort  in  theory to  interpret  its  data. Nowadays,  detectors  and
computers  are cheap  and powerful,  providing hardware  solutions for
FASS, FADE,  RINGSS, etc.  The  emphasis is  on the theory  needed for
correct  interpretation  of  their   signals  and  on  the  associated
software.

\section*{Acknowledgments}

This work was  performed at the NSF's NOIRLab  and partially supported
by the  award 1421197 from the NSF.  It is stimulated by  the plans to
test  new  astronomical  sites.  I   am  grateful  to  the  FASS  team
(A.~Guesalaga  and B.~Ayanc\'an) for  helpful discussions  of turbulence
profiling,  sharing  their results,  and  pertinent  comments on  this
study.    Comments  by  the  anonymous   Referee  are  gratefully
  acknowledged.

\section*{Data Availability}

No data  were generated in  this research. The  IDL programs used  in this
work  is  an experimental  software  which  is  not yet ready  for  public
distribution.




\bsp

\label{lastpage}

\end{document}